\documentclass[twocolumn,linenumbers]{aastex631}

\usepackage{amsmath,amstext}
\usepackage{mathtools}
\hypersetup{
    colorlinks=true,
    linkcolor=blue,
    filecolor=magenta,      
    urlcolor=cyan,
}

\shorttitle{}
\shortauthors{Jaehnig, Bird, and Holley-Bockelmann}

\newcommand{\eg}{{\rm e.g.}}

\newcommand{\ie}{{\rm i.e.}}

\newcommand{\pmem}{\ensuremath{P_{\mathrm{mem}}}}
\newcommand{\ra}{\ensuremath{\alpha}}
\newcommand{\dec}{\ensuremath{\delta}}
\newcommand{\pmra}{\ensuremath{\mu_{\alpha*}}}
\newcommand{\pmdec}{\ensuremath{\mu_{\delta}}}
\newcommand{\plx}{\ensuremath{\varpi}}

\newcommand{\spmra}{\ensuremath{\sigma_{\mu_{\alpha*}}}}
\newcommand{\spmdec}{\ensuremath{\sigma_{\mu_{\delta}}}}
\newcommand{\splx}{\ensuremath{\sigma_{\varpi}}}

\newcommand{\cplxpmra}{\ensuremath{\rho_{\varpi \mu_{\alpha*}}}}
\newcommand{\cplxpmde}{\ensuremath{\rho_{\varpi \mu_{\delta}}}}
\newcommand{\cpmrapmde}{\ensuremath{\rho_{\mu_{\alpha} \mu_{\delta}}}}

\newcommand{\g}{\ensuremath{G}}
\newcommand{\bp}{\ensuremath{G_{BP}}}
\newcommand{\rp}{\ensuremath{G_{RP}}}
\newcommand{\normal}[2]{\ensuremath{\mathcal{N}(#1, #2)}}
\shorttitle{Memberships using XDGMM for 431 open clusters}
\shortauthors{Jaehnig et al.}
\begin{document}
\defcitealias{Cantat-Gaudin2018b}{CG2018b}

\title{Membership lists for 431 open clusters in Gaia DR2 using extreme deconvolution gaussian mixture models}

\author[0000-0002-7916-1493]{Karl Jaehnig}
\affiliation{Department of Physics and Astronomy, Vanderbilt University, Nashville, TN 37235, USA}
\affiliation{Fisk-Vanderbilt Bridge Fellow}
\affiliation{LSSTC Data Science Fellow}

\author{Jonathan Bird}
\affiliation{Department of Physics and Astronomy, Vanderbilt University, Nashville, TN 37235, USA}
\affiliation{VIDA Fellow}

\author[0000-0003-2227-1322]{Kelly Holley-Bockelmann}
\affiliation{Department of Physics and Astronomy, Vanderbilt University, Nashville, TN 37235, USA}

\affiliation{Department of Physics, Fisk University, Nashville, TN, 37208, USA}


\begin{abstract}

Open clusters are groups of stars that form at the same time, making them an ideal laboratory to test theories of star formation, stellar evolution, and dynamics in the Milky Way disk. However, the utility of an open cluster can be limited by the accuracy and completeness of its known members. 
%We develop an automated method to extract known open clusters from Gaia DR2 data employing extreme deconvolution gaussian mixture models, taking into account all of the astrometric uncertainties, correlations, and covariances.
%We extract known open clusters from Gaia DR2 data in a ``top-down'' fashion, 
Here, we employ a ``top-down'' technique, {\it extreme deconvolution gaussian mixture models} (XDGMM), to extract and evaluate known open clusters from Gaia DR2 by fitting the distribution of stellar parallax and proper motion along a line-of-sight.
Extreme deconvolution techniques can recover the intrinsic distribution of astrometric quantities, accounting for the full covariance matrix of the errors; this allows open cluster members to be identified even when presented with relatively uncertain measurement data. To date, open cluster studies have only applied extreme deconvolution to specialized searches for individual systems.
%Extreme deconvolution techniques incorporate even noisy data while accounting for astrometric measurement uncertainties, correlations, and covariances, and have thus far only been applied in specialized searches for individual OCs.  
We use XDGMM to characterize the open clusters reported by \citet{Ahumada2007} and are able to recover 420 of the 426 open clusters therein (98.1\%). Our membership list contains the overwhelming majority ($>95\%$) of previously known cluster members. We also identify a new, significant, and relatively faint cluster member population and validate their membership status using Gaia eDR3.
%newlyoverlaps with 96.1$\%$ of the membership lists in the literature, and we find cluster stars not in the literature that are on average 1.1 magnitudes fainter.
% {\bf sentence about validation of membership list and the fact that you've uncovered a suite of dimmer/noisier members} %\bf{We further confirm the existence of 11 known open clusters within Gaia Dr2.} \jon{Leaning towards removing preceeding sentence}.
%We report 
We report the fortuitous discovery of 11 new open cluster candidates within the lines of sight we analyzed. We present our technique, its advantages and challenges, as well as publish our membership lists and updated cluster parameters. 
\end{abstract}

\keywords{Methods: data analysis -- open clusters: general -- astrometry} 

\section{Introduction} \label{sec:intro}
Open clusters (OCs) are single stellar populations that formed together. Thousands of OCs have been identified and are widespread throughout the Milky Way disk \citep{Friel1995, Kharchenko2013, Dias2014}.
%and are a well-known birthplace of stars \citep{Larson1995}. 
Studying OCs offers insights into theories of star formation  \citep{Kroupa2001} and provides a direct view of the high mass end of the stellar initial mass function. OCs are also sites of few-body dynamical mechanisms that can produce unique astronomical objects such as Blue Straggler stars \citep{Perets2015, Toonen2020ThePathways, He2017a}. Taken as a whole, the OC population are key tracers of chemical enrichment throughout the Galactic disk \citep{Spina2021}.

\par 

\begin{figure}
%\hspace*{-1.5cm}
% \epsscale{1.2}
\begin{center}
\includegraphics[scale=0.75]{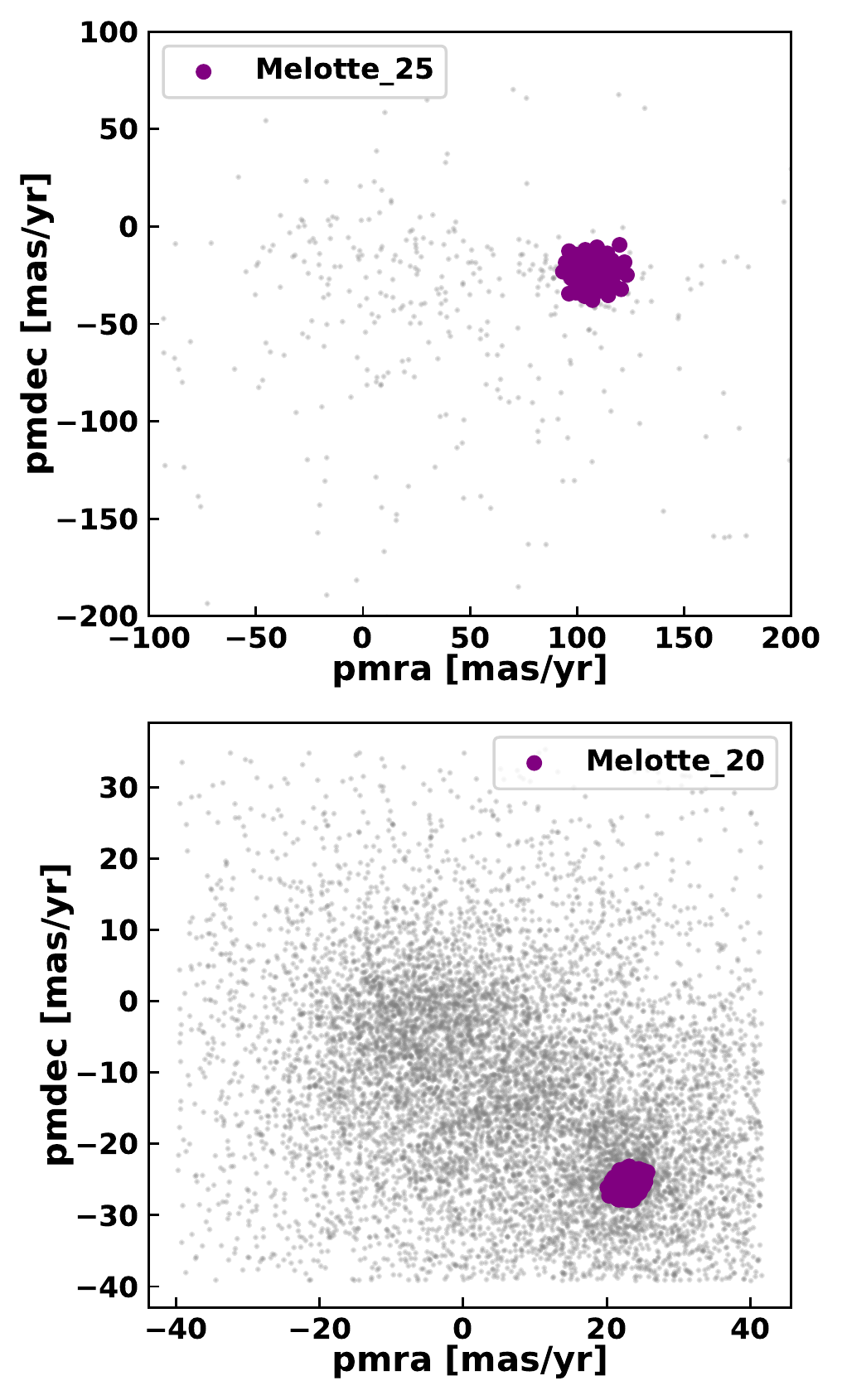}
\parbox{\columnwidth}{\caption{Proper-motion vector plot for Melotte-25, aka The Hyades Cluster (top panel, purple dots), and Melotte-20, aka the Alpha Persei Cluster (bottom panel, purple dots). Manually selecting open cluster members using vector point diagrams in proper-motion are easier (top panel) for some clusters than for more embedded clusters (bottom panel). \label{fig:twoplot} }} 
\end{center}
\end{figure}

Identifying the members of a particular open cluster can be difficult and numerous techniques have been developed to tackle the problem.  Many techniques rely on finding the population of stars in proper-motion space. In proper-motion space, the population of stars that form an OC will appear as a compact group of similar proper-motions. This grouping can be identified straightforwardly for those open clusters whose proper-motions are high enough that they are separate and distinct from the galactic field stars (\eg, Melotte 25 (the Hyades) in Figure \ref{fig:twoplot}, top-panel). 
%\par 
However, more embedded OCs, such as Melotte 20 (see Figure \ref{fig:twoplot}, bottom-panel), present a greater challenge.

\par
With the advent of precise astrometric data from the Gaia survey \citep[see][]{Gaia_mission2016, Gaiadr2_2018}, it is now possible to distinguish and computationally extract open clusters even when the kinematics of the cluster and field stars are not significantly different from one another. We briefly review some commonly employed methods here, but see \citet{Hunt2020ImprovingData} for a thorough and comprehensive review.

Cluster and cluster member identification algorithms that use astrometric data leverage the compact nature of a cluster in some combination of position, distance, and proper motion. Most techniques employ what could be called a `bottom-up' approach in which the star cluster is built up out of smaller components until all the stars within a line of sight are either considered a 'cluster' star or a 'field' star.

Algorithms such as K-means~\citep{Utsunomiya1996}, fit a K number of clusters to data by iteratively finding K mean centroids until some convergence threshold is met. Other algorithms, such as DBSCAN~\citep{Daszykowski2009}, assign groupings of stars to clusters using inter-particle distance. One benefit of these algorithms is that they are not confined to proper-motion data, and have been successfully applied to position and multi-dimensional astrometric data~\citep{Gao2018,Castro-Ginard2018}.  Both DBSCAN and K-means have been used to identify stars in known open clusters, but they have also been successfully employed to find new open cluster candidates \citep{Cantat-Gaudin2019GaiaPerseus, Liu2019a, Castro-Ginard2019}.

\par 

Despite its power and versatility, K-means algorithms necessitate pre-defining the number of clusters that are to be found. This may be difficult to do without explicit prior knowledge of
how the data are distributed along the dimensions of interest. Workflows using K-means also typically assume that the data are all distributed spherically, a clear limitation when presented with more realistic cluster shapes \citep{Pang20213DDynamics}.
%\par
DBSCAN is more generalized than K-means in that it can find non-symmetric clusters distributed throughout the data, and also doesn't require a preset number of clusters. However, it becomes computationally expensive when considering large data sets in more than 3 dimensions, as all of the pairwise distances between data points must be calculated. The hyperparameters used in DBSCAN to define clusters also needs to be fine-tuned, requiring hyperparameter optimization which can become computationally expensive.

\par 
Once OC candidates have been selected and preliminary membership is assigned, it is still necessary to ensure that the selected group of stars is a true physical cluster. %effectively selected from the raw data correctly. 
One of the more reliable confirmation techniques for the overall cluster is to directly inspect the candidates in color-magnitude space. 
A group of stars with a narrow range of parallax can masquerade as a cluster within a CMD.
However, false cluster members with erroneous parallax measurements will appear in unphysical locations within the CMD. Requiring member stars to be clustered in proper-motion as well as distance also limits false-positive cluster identification.
As open clusters represent a single stellar population, the color-magnitude plot will exhibit well-known properties: the main sequence, the binary main sequence, and the red giant branch sequence if the cluster is sufficiently old.
This technique becomes even more effective when previous values of the star cluster are known, such as metallicity ([Fe/H]), age, and extinction (A$_{V}$). A theoretical isochrone can then be generated to compare the observed colors and magnitudes of the stars with what would be expected from stellar evolutionary theory.
\par 
Perhaps the hardest aspect of such confirmation methods is that they are often performed manually for each individual OC. This becomes a daunting task, considering that the Milky Way Star Cluster (MWSC) catalog \citet{Kharchenko2013} contains over 2000 open clusters. 

There have been great strides in automating the process of open cluster membership construction to be able to take a census of the Milky Way population. 

\citet{Cantat-Gaudin2018b} recently performed one of the largest OC automated analyses in Gaia DR2, deriving mean cluster parameters (\ra, \dec, \pmra, \pmdec, \plx, $r_{core}$, etc) as well as individual membership lists for 1229 OCs throughout the Milky Way. \citet{Cantat-Gaudin2019} updated these results and grew their catalog to 1481 clusters.  

Their method relies on finding all sub-clumps defined by running a K-means algorithm in their chosen sight lines. They then use the minimum spanning tree metric, $\Lambda$ \citep{Allison2009}, to distinguish between concentrated star cluster sub-clumps and uniformly distributed sub-clumps of the same number that are found in the field. This was done iteratively to construct their final open clusters out of these sub-clumps. Finally, they produce membership probabilities by re-sampling the 3x3 covariance matrices of the stars and re-running their algorithm.

\par 
'Top-down' cluster identification algorithms, which aim to characterize the environment that includes the cluster, offer an alternative to the 'bottom-up' approach.
Gaussian Mixture Models (GMMs) \citep{Pearson1893ContributionsEvolution} describe a data set using a number of multivariate Gaussian components. GMM is a 'top-down' method that describes a line of sight (or subset thereof) as a mixture of multivariate gaussians. An individual gaussian, or component, of the model represents the cluster. GMMs have already been used to identify and characterize a number of astronomical objects, including individual open clusters \citep{Gao2018}, the galaxy red sequence \citep{Hao2009}, and supernovae/host galaxy populations \citep{Holoien2016}.

\par 
GMMs allow for mixed memberships, wherein a data point can have a probability of belonging to multiple components. K-means and DBSCAN typically assign each data point only to one component or cluster. GMMs also have fewer hyperparameters to tune than DBSCAN or other similar agglomerative clustering methods. The resulting covariance matrices from a well-fit GMM effectively describe the shape, scaling, and orientation of components within the data.

\par 
As \citet{Cabrera-Cano1990AClusters} cautions in their analysis of $\sim$ 300 stars in NGC-2420, care must be taken when fitting an OC dataset with GMMs. If the field star fraction is too large, the OC may be overwhelmed by field stars along the line of sight. An open cluster may be also be less separable if the distribution of field stars is strongly non-gaussian. Both of these situations may produce less than optimum results, although a quantitative analysis of the effect of these potential issues on a given cluster was not addressed.

It is ideal for a GMM to fit an intrinsic, underlying distribution rather than noisy discrete data. To accomplish this the observed measurements must first be deconvolved from their uncertainties  \citet{Bovy2011} developed a method to determine the underlying distribution function in the presence of noisy data called `Extreme Deconvolution'.

Extreme Deconvolution Gaussian Mixture Models (hereafter, XDGMM) combine the two techniques to ideally fit the true, intrinsic distribution with a gaussian mixture model. XDGMM has already been employed successfully to characterize members of a Galactic open cluster \citep{Olivares2019RuprechtDANCe}  and to calculate mean cluster positions and proper-motions in the case of an actively disrupting open cluster \citep{Price-Whelan2018}. However, these use cases of XDGMM were for a small number of individual systems.

In this paper, we use XDGMM to identify members of known open clusters within Gaia DR2. We construct an automated pipeline to address the shortcomings of GMM cluster fitting -- computational cost, the number of gaussian components to fit, and selection of the gaussian component associated with the cluster.

In Section \ref{sec:data} we discuss the Gaia DR2 data. In Section \ref{sec:methods} we describe our method to extract open clusters. We present the mean cluster parameters and the properties of designated cluster members as well as a validation of our results via comparison with the literature in Section \ref{sec:results}.  In Section \ref{sec:serendipitous_ocs}, we describe the serendipitous discovery of 11 new candidate open clusters not found in the current literature. 
%We discuss some peculiar cases of XDGMM fits in Section \ref{sec:discussion}.
Finally,  we summarize our findings on open cluster extraction using an automated machine learning method and discuss future work in Section \ref{sec:summary}.
%{\bf downloaded}

\section{Data} \label{sec:data}

We first assemble a list of open clusters to consider using the  cluster catalog 
from \citet{Ahumada2007}, originally created in their search for blue straggler stars. They constructed a catalog of 427 open clusters throughout the Milky Way, using Johnson UBV photometry~\citep{Johnson1955} and isochrones~\citep{Girardi2000Evolutionary0.03}.
We remove one known asterism \citep[NGC-1252][]{Kos2018TheClusters, Angelo2019, Cantat-Gaudin2019} from consideration.
This reduces the total number of open clusters we consider to 426.

The effectiveness of any search for or characterization of open clusters will depend upon the input data.
Gaia DR2 contains $5$ dimensional astrometric measurements -- position (\ra, \dec), proper motion (\pmra, \pmdec), and parallax (\plx) -- and photometry for over $1.3 \times 10^9$ sources \citep{lindegren2018astrometric}.
The quality of this data set makes it ideal for open cluster searches while its quantity presents a challenge.
Attempting a global XDGMM fit over the entire sky is computationally prohibitive and would yield a suboptimal result even if possible (Section~\ref{sec:methods}).
 
For each cluster, we define a target field of potentially associated Gaia DR2 sources.
The target field of view is centered on the position of the cluster and has conservatively-wide opening angle to ensure that we do not remove potential cluster members and affect our subsequent analysis.
While \citet{Ahumada2007} report both cluster positions and angular sizes, we supplement their measurements with the more recent open cluster catalog efforts of \citet{Dias2018} and \citet{Kharchenko2013}.
The addition of these catalogs and variation amongst them also captures the uncertainty of these input cluster parameters in our analysis.
Our target field center is the median of the positions reported in these cluster catalogs.
The opening angle is equal to $1.5\times$ the largest cluster angular diameter reported amongst  \citet{Ahumada2007}, \citet{Dias2018}, and \citet{Kharchenko2013}.

We searched for all Gaia DR2 sources within the 426 target fields of view. 
We obtained the astrometric [\ra, \dec, \pmra, \pmdec, \plx] and photometric [\g, \bp, \rp] measurements, as well as the relevant uncertainties and correlation coefficients.
In addition to the Gaia DR2 data, we collected all previous distance determinations to the clusters in our sample through the  \texttt{SIMBAD}\citep{Wenger2000} database.
These distance measurements will be used in a pre-processing step described in Subsection~\ref{ssec:data_prep}.
We used the \texttt{ASTROQUERY} python module \citep{Ginsburg2019Astroquery:Python} to query both Gaia DR2 and \texttt{SIMBAD}.

\section{Methods} \label{sec:methods}
Ultimately, we identify a cluster and its members via a XDGMM fit of the density field represented by standardized \pmra, \pmdec, and \plx\ measurements and uncertainties of stars in the cluster vicinity.
Our procedure's computational cost and its effectiveness correlates with the density contrast between the cluster and field. 
We address both of these concerns by first restricting each target field to a manageable subsample. 

The position information of the subsample of stars associate with each cluster is only used for validation purposes (Section~\ref{sssec:visual_validate}). 
We make this decision based on the work of \citet{Hillenbrand1998ADynamics}, \citet{Kuhn2014TheSubclusters}, and \citet{Kuhn2017b}. These works considered clustering of star clusters using their observed on-sky positions. They found that a multivariate normal distribution may not be best fit to these distributions. 

We scale the measurements and uncertainties of interest to account for the different variance of each feature before applying the XDGMM fit. Finally, we establish quantitative metrics to select the number of components to include in the XDGMM fit and to select the individual component corresponding to the cluster. 
The details of our method follow.

%\subsection{Preparing the data} \label{ssec:data_prep}
\subsection{Data Preprocessing} \label{ssec:data_prep}

The Gaia DR2 data is known to have systematics that, if unaccounted for, can lead to erroneous or biased results~\citep{Vasiliev2019SystematicClusters}. We use the recommendations from \citet{lindegren2018astrometric} to account for systematics found within the parallax and proper-motion data. 
In particular, we subtract the -0.029 mas parallax zero-point from all of our stars and exclude all stars fainter than $18^{th}$ Gaia G magnitude.
Our faint limit reduces the total number of stars included in our analysis. Stars with $m_{\mathrm{G}} > 18$ typically have high fractional astrometric uncertaines in Gaia DR2 making them ill-suited for cluster-finding \citep[\eg][]{Cantat-Gaudin2019, Hunt2020ImprovingData}.
We also correct for the inertial spin found within the proper-motions of bright sources\footnote{Gaia DR2 $G$ $< 11$} using the recommendations from \citet[][]{Lindegren2019TheStars}. 
% \joncomment{How is the correction done?}

After applying the global corrections and limits, we preprocess the input data within each cluster target field independently.
The goal is to remove obvious field stars and astrometric outliers from consideration as cluster members.
This reduces the dynamic range of parallax and proper motion and, in turn, increases the signal of any over-density associated with the cluster.

We first remove outlier stars in proper motion.
Stars with \pmra\ or \pmdec\ measurements more than $10\sigma$ away from the median of their target are dropped from consideration.
Individual high proper-motion stars may skew the feature scaling within a target field.
This step is performed for all clusters except for the high proper-motion cluster Melotte-25 (the Hyades cluster).

We remove parallax outliers only if more than $10^4$ stars in a target field remain eligible for XDGMM fitting.
We found that this numerical threshold strikes a good balance between ensuring a complete cluster membership list given our magnitude limit and maintaining computational efficiency of the XDGMM fitting. 
If a target sample contains $>10^4$ stars, we discard the stars least likely to be at the distance of the known cluster.

The open clusters have uncertain literature distances and there is uncertainty measured parallax of each star.
For these reasons, we do not simply discard stars using a strict delta parallax cut-off.
Rather, we employ a simple probabilistic model to determine the stars least likely to be within the literature parallax range of the cluster.

%The following parallax truncation is only performed on LOSs which have more than 1e4 stars within the dataset up to this point.
We assume the parallax of cluster members is distributed normally with a mean $\mu$ and standard deviation $\sigma$ given by:
\begin{equation}\label{equ:cluster_plx_dist}
    \begin{split}
    \mu=med(\plx_{lit}), \\
    \sigma=
    \begin{cases}
    0.25\cdot med(\plx_{lit})\ ,& \text{if } med(\plx_{lit})> 1\ mas\\
    0.25\ mas, & \text{if } med(\plx_{lit})< 1\ mas\\
    \end{cases}
    \end{split}
\end{equation}

where $med(\plx_{lit})$ is the median of the cluster parallaxes reported in the three cluster catalogs sources we examine (Section ~\ref{sec:data}). We set $\sigma$ to depend on $med(\plx_{lit})$. Clusters with a literature parallax less than 1~mas (distance greater than 1~kpc) have a constant $\sigma_{\varpi}$ of 0.25~mas. Clusters with a literature parallax greater than 1~mas have their $\sigma_{\varpi}$ set to a 25\% of their median literature parallax. This ensures that the parallax distribution of nearby clusters is not excessively truncated.

\par 
We use bootstrap sampling to calculate the probability that each stellar parallax measurement is drawn from the cluster parallax distribution. For each star, we sample a normal distribution with mean $\mu=\plx_{obs}$ (\textit{gaiadr2.parallax}) and $\sigma=\sigma_{\plx_{obs}}$ (\textit{gaiadr2.parallax\_error}) 1000 times. We then evaluate the PDF of the corresponding cluster, defined in equation \ref{equ:cluster_plx_dist}, at the locations of these 1000 samples. The star's probability of parallax overlap is then the mean of the 1000 PDF evaluations.

Only the $10^4$ stars with the highest parallax overlap probability within the target field are kept and subject to XDGMM fitting.
After this step, all target fields have been reduced to a maximum of $10^4$ stars based on the closeness of their parallax and proper motion to the parent cluster.
All of the analysis and XDGMM fit procedure described below pertain only to these restricted data.

\subsection{Scaling the \pmra, \pmdec, \plx\ data}\label{ssec:scaling}
\par 
Feature scaling is necessary to establish an appropriate shape of the parameter space as we are fitting 3-dimensional data with unequal means and variance.
We use 'StandardScaler' from the \texttt{Scikitlearn} \citep{Pedregosa2015Scikit-learn} python package, which shifts and scales the data in each to dimension to a mean of zero and a variance of one.

We first smooth the data before scaling to ensure we calculate the scaling parameters accurately.
The smoothed distribution is a simple resampling of the [\pmra, \pmdec, \plx] distribution according to the uncertainty covariance matrix (equation \ref{equ:covar_matrix}). 
We resample each star N$_{samples}$ times such that StandardScaler uses approximately $5\times10^4$ coordinate positions (\ie, N$_{samples} \times$ N$_{data}$ = $\sim5\times10^4$, where N$_{data}$ is the number of stars remaining in the field after pre-processing).
The shift and scale parameters found for the smooth distribution are then applied to the actual measurements prior to fitting.
We scale the measurement uncertainties as well, to preserve the fractional uncertainty of each measurement in the transformed data space.

%\subsection{Fitting the LOS with XDGMM and cluster identification}
\subsection{XDGMM fit of the scaled data}
\label{ssec:xdgmm_fit}

Gaussian Mixture Models employ the assumption that any N-dimensional data can be represented as the summation of K components, each of which is an N-dimensional Gaussian. The K Gaussian distributions have K means, and K covariances. 

XDGMM takes into full account the full covariance matrix representing the uncertainty on \pmra, \pmdec, and \plx, and the correlation $\rho$ for each data point. This covariance matrix is assembled as written in equation \ref{equ:covar_matrix}.
\vspace{2mm}
\begin{equation}\label{equ:covar_matrix}
\vspace{5mm}
\begin{pmatrix*}
    \spmra^2 & \spmra \spmdec \cpmrapmde & \spmra \splx \cplxpmra \\ \cr
    \spmdec \spmra \cpmrapmde & \spmdec^2 & \spmdec \splx \cplxpmde \\ \cr
    \splx \spmra \cplxpmra & \splx \spmdec \cplxpmde & \splx^2 
\end{pmatrix*}
\end{equation}

We perform an XDGMM fit of the scaled astrometric data and the associated scaled covariance matrices using the XDGMM python wrapper \citep{Holoien2016}.
Each cluster target field is fit independently.

\subsubsection{Determining the number of XDGMM components}
\label{sssec:xdgmm_ncomp}

We fit each cluster-specific scaled data set nine times, using from 2 to 10 Gaussian components. We do not fit more than 10 components due to the  computational cost. All model fits converge within a tolerance of $1\times10^{-8}$ and use a random seed of 999.

Of the nine XDGMM fits to a cluster, we select the best-fit model according to the Bayesian Information Criterion \citep[BIC][]{GhoshJayantaK.2006AnMethods}. The BIC is commonly used to select between models when fitting to a data set as it strikes a balance between maximizing the likelihood and penalizing the introduction of new parameters to avoid over-fitting.
The BIC is defined as,

\begin{equation}
    {\rm BIC}\ =\ x\cdot \log{n}\ -\ log{\hat{L}},
\end{equation}
where x is the number of parameters needed in the XDGMM fit, n is the number of data points, and $\hat{L}$ is the maximized likelihood of the model.
The best-fit model has the lowest BIC score.

We find that the optimal number of components in the best-fit model is typically between 5 and 8 and correlates with the number of stars in the target field (Figure~\ref{fig:k_comp_hist}).
Heavily populated target fields (\textgreater 5000 stars; black dashed histogram) generally required more components in the best-fit model than sparsely populated fields (red histogram).

\subsubsection{Finding the Open Cluster component} \label{sssec:cluster_selection}

We use the differential entropy information metric to select the gaussian component that best describes the open cluster. 
Differential entropy is a measure of how compact a distribution is within a volume. In this case, we can ascertain the compactness of each gaussian component in the scaled data space. 
For a multivariate gaussian, the differential entropy ($h$) can be written as,

\begin{equation}\label{equ:diff_entropy}
    {\it{h}} = \frac{d}{2} + \frac{d}{2}ln(2\pi) + \frac{1}{2}ln(\bold{|\Sigma|}),
\end{equation}

where d is the dimension of the parameter space and $\mathbf{|\Sigma|}$ is the determinant of the covariance matrix for the multivariate gaussian of each component in the best-fit XDGMM \citep[see][]{Ahmed1989EntropyDistributions}.

We automatically designate the component with the lowest differential entropy as the cluster component.

\subsubsection{Individual membership probabilities}\label{sssec:memb_proba}

Once we have determined the optimal number of components and which component belongs to the open cluster, we employ bootstrap resampling to calculate the open cluster membership probabilities of each star in the scaled target field. We assign each star within a target field to a component using the XDGMM code's built-in 'predict' function. We then generate a new scaled [\plx, \pmra, \pmdec] value for each star from its scaled 3x3 covariance matrix. We recompute component assignments for each star for a total of 100 iterations. The final membership probability of any one star is the number of times that the star was assigned to the open cluster component out of 100.

\subsubsection{Validation of fit results}\label{sssec:visual_validate}
\par
\begin{figure}
\hspace*{-1.25cm}
% \epsscale{1.2}
% \begin{center}
\includegraphics[scale=0.55]{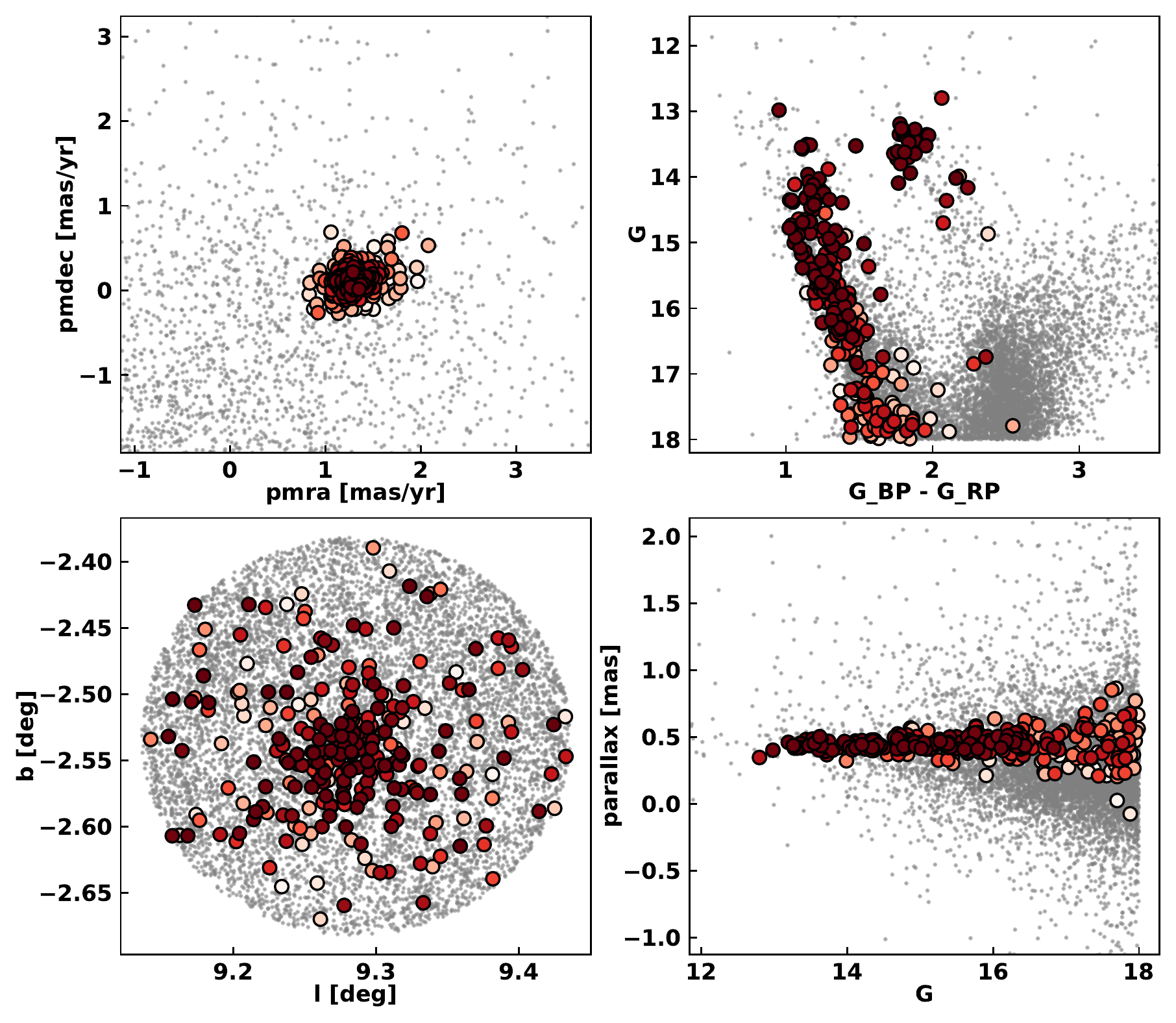}
\parbox{\columnwidth}{\caption{An example four panel plot used for visual validation of a fully automatic XDGMM extraction. The open cluster is NGC-6583. In all four panels, the field is given by gray dots, and the cluster stars are given by the red circles, where the lower membership probability stars tend towards white in color. The top left panel is a vector point diagram plot in proper motion. The top right panel is a color-magnitude diagram using Gaia filters. The bottom left panel is a plot of positions in galactic coordinates. The bottom right panel is a plot of the parallax measurements for cluster stars as a function of their Gaia G-band magnitude. \label{fig:ngc6583_four_plot}}}
% \end{center}
\end{figure}
\begin{figure}[ht!]
% \hspace*{-.5cm}
% \epsscale{1.2}
% \begin{center}
\hspace*{-.5cm}
\includegraphics[scale=0.6]{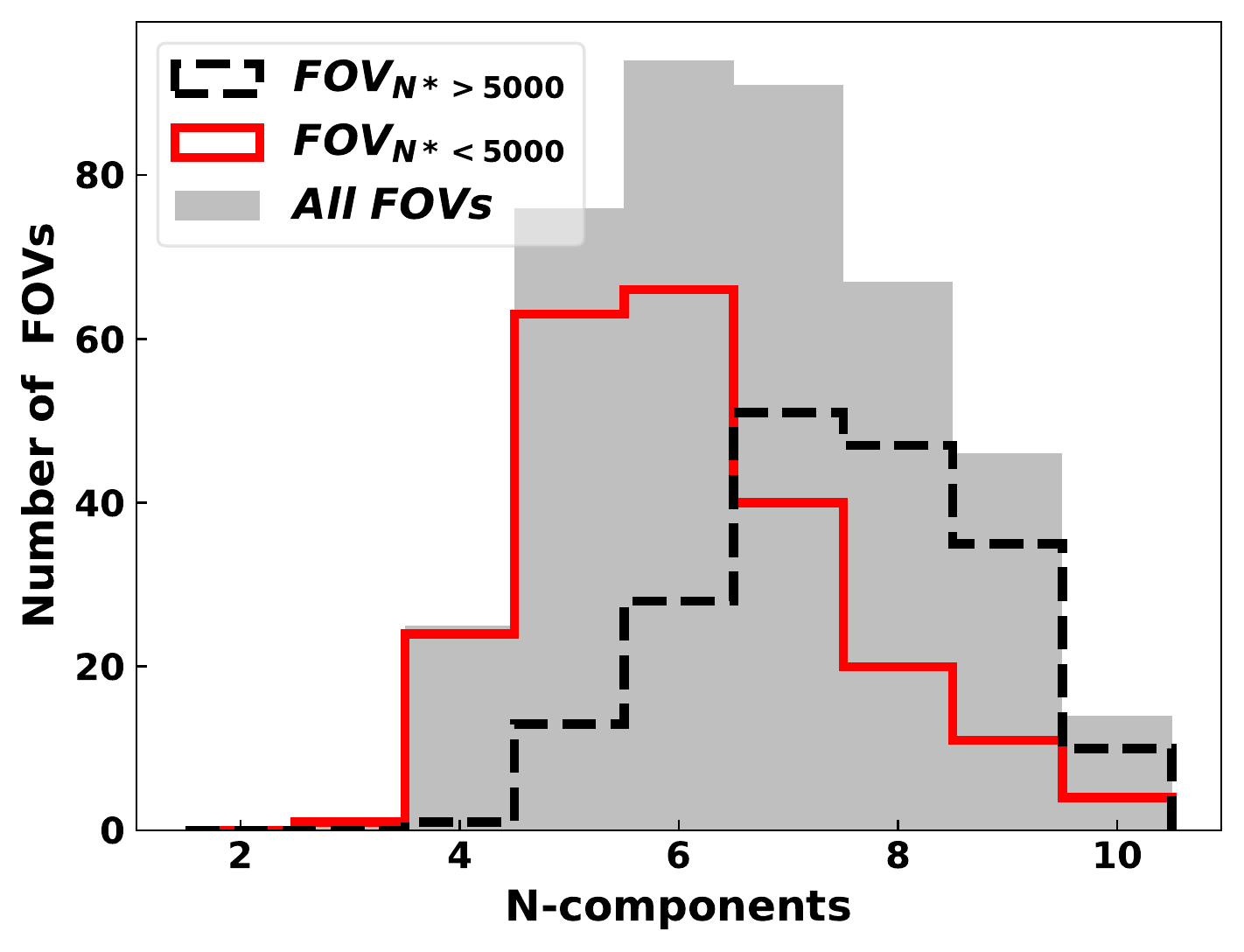}
\parbox{\columnwidth}{\caption{Histogram of the number of components selected successfully with the BIC for 414 target field XDGMM fits. The complete distribution is given by the gray shaded histogram. The red solid line is the distribution of the number of components for all target fields with less than $5000$ stars. The dashed black line is the distribution of the number of components for all target fields with more than $5000$ stars. We find that the more populated target fields required more components to disentangle the open cluster from the field than sparser target fields. \label{fig:k_comp_hist}  }} 
% \end{center}
\end{figure}

We visually inspect the resulting open cluster model results by plotting four different characteristics of the cluster stars with a membership probability above 50\%. We plot the proper-motion of the stars in \ra\ and \dec\, the \g\ and \bp\ - \rp\ color-magnitude diagrams, the positions in {\it{l}} and {\it{b}}, and \plx\ as a function of G-band magnitude. %This format is similar to Figure 5 in \citet{Cantat-Gaudin2018a}.
We show these four panels for the cluster NGC-6583 in Figure \ref{fig:ngc6583_four_plot} as an example.

We define a successful open cluster retrival, in the broadest sense, via visual validation. Statistical tests can be found in Section~\ref{sec:results}.
We retrieve a known open cluster when the resulting members (probability \textgreater 50\%) are concentrated in proper-motion vector space, form a relatively well-defined isochronal color-magnitude sequence, are centered within the target field on-sky, and occupy a narrow distribution in parallax as a function of G-band magnitude.

\section{Results} \label{sec:results}

We successfully recover 420 of the 426 known open clusters targeted. 
The probable members of these identified clusters are highly concentrated in proper-motion and parallax space relative to field stars and populate a narrow isochrone in the color-magnitude diagram (Section \ref{sssec:visual_validate}). 
We statistically validate our cluster membership designations and the astrometic location of the recovered clusters in Section~\ref{ssec:res_validate}.

Our automated pipeline to select the best-fit XDGMM and identify the individual Gaussian component corresponding to the cluster target required no human intervention in $95\%$ of the investigated target fields.
In these $385$ fields, the lowest differential entropy ($h$) component of the best-fit model represented the targeted cluster.
The remaining $41$ fields required manual analysis of XDGMM fits.
We found the target cluster was fit by the second lowest $h$ component of the best-fit model in $29$ of the manual analysis fields.
In nearly all ($28/29$) of these target fields, the minimum $h$ component was either a spatially adjacent known open cluster or a new open cluster candidate (Section~\ref{sec:serendipitous_ocs}).
All told, the minimum $h$ component of the best-fit XDGMM identified a cluster or candidate cluster in $413$ of the 426 target fields.

None of the best-fit model components fit describe the known cluster in remaining $12$ target fields.
We visually inspected all 
54 ($\sum_{i=2}^{10} i$) components within the 9 BIC-scored model fits to each of these target fields in search of a component that corresponded to the known open cluster.
We identified a Gaussian component that represented the open cluster in 6 of these target fields.

In the remaining 6 target fields, we were unable to recover a cluster candidate component either automatically or manually despite convergence of the XDGMM code. At least 2 of these 6 targeted clusters likely violate our assumption of the open cluster being a significant over-density in the [\pmra, \pmdec, \plx] space. Ruprecht-46 may be an asterism masquerading as a cluster, as was recently argued by \citet{Cantat-Gaudin2020}. Collinder-228 is a well-known cluster in the Carina Nebula, but \citet{Feigelson2011X-rayComplex} found that the cluster is a composition of many sparse groups with no clear central concentration.

Nevertheless, our automated pipeline was successful in a variety of conditions.
Within the 385 target fields with a fully automated and successful target cluster recovery:

\begin{enumerate}
    \item The target field contains a total number of stars as few as 123 stars and as many as $1\times10^4$ stars
    \item The number of cluster member stars (with a membership probability $\textgreater$ 0.5) was as few as $40$ stars and as many as $\sim$ 3700 stars
    \item Cluster members comprised a wide range, $0.64\%$ -- $54.5\%$, of all stars in the target field
\end{enumerate}

This work is, to the best of our understanding, the first to report central parameters using Gaia DR2 for some of the target clusters.
We identify 11 clusters that are not included in the Gaia DR2 based cluster catalogs of \citet{Cantat-Gaudin2018b}, \citet{Cantat-Gaudin2019GaiaPerseus}, \citet{Castro-Ginard2018}, \citet{Castro-Ginard2019} \citet{Castro-Ginard2020HuntingDisc},  \citet{Ferreira2020DiscoveryDR2}, \citet{Sim2019}, \citet{Liu2019}, or \citet{Hunt2020ImprovingData}. Cluster parameters for these systems can be found in  Table~\ref{tab:cluster_mu_params}.

Our cluster member catalog contains $261\,662$ stars with non-zero membership probability (\pmem). 
In our subsequent analysis, we define members, probable members, and highly probable members of a cluster as those stars with $\pmem$ greater than $0.5$, $0.85$, and $0.95$, respectively.

The member catalog of the 420 recovered open clusters includes $171\,939$ cluster members, $125\,235$ probable members, and $95\,920$ highly probable members. We list 10 entries for 10 cluster members and their Gaia DR2 measurements in Table~\ref{tab:membs_params}.

Finally, we report the serendipitous recovery of several candidate open clusters. The XDGMM component corresponding to the candidate clusters had low differential entropy and the properties of the candidate members are consistent with open clusters. These candidates were discovered within the 426 target fields but are not found in the \citet{Ahumada2007} catalog. We describe our method for identifying these open cluster candidates and their properties in Section~\ref{sec:serendipitous_ocs}.

\subsection{Cluster Astrometric Parameters and Distances} \label{ssec:cluster_param}

We determine robust central astrometric coordinates [\ra, \dec, \plx, \pmra, \pmdec] of each cluster using the measurements of probable members (Table~\ref{tab:cluster_mu_params}).

To do so, we first remove potential outlier measurements via a sigma-clipping procedure.
We calculate an initial coordinate median and median absolute deviation (MAD), both weighted by inverse measurement uncertainty, of probable members.
We estimate $\sigma$ assuming a normal distribution using $\hat{\sigma}=1.4826 \times \mathrm{MAD}$ \citep[\eg,][]{huber1980}.
We discard all outlier measurements not within $\pm$3 $\sigma$ of the initial weighted median.

We then recalculate the weighted median and MAD of the non-outlier stars for each of the 5 astrometric parameters. 
%The final weighted MAD statistic is multiplied by k=1.4826 to approximate a normal distribution standard deviation. 
We list these median values and the standard deviation estimator $\hat{\sigma}$ in Table \ref{tab:cluster_mu_params}. We also calculate the r$_{50}$ radius, containing half of the cluster members, using the cluster positions and list these values in \ref{tab:cluster_mu_params}.

Cluster distance, while not a direct astrometric measurement, can be ascertained to a relatively high precision 
%than many individual stars 
because each cluster member can be assumed to lie at nearly the same distance along the line of sight. 
Simple parallax inversion is not adequate to measure physical distances using Gaia DR2.
% \joncomment{do you mean a simple 1/parallax or something else?} 
This is due to the nature of the Gaia DR2 parallax measurements which may be close to zero or even negative. We obtain individual cluster distances through Bayesian Inference \citep[see][for more details]{Luri2018}.

The posterior probability of the cluster distance can be expressed as,

\begin{equation}\label{equ:dist_posti}
    P\big (r |\{\plx\},\{\sigma_{\plx}\},L\big) \propto P\big(r|L\big)P\big(\{\plx\}|r,\{\sigma_{\plx}\}\big),
\end{equation}
where $P(r|L)$ is the exponential decreasing space density prior~\citep{Bailer-Jones2015EstimatingParallaxes}:
\begin{equation}\label{equ:dist_prior}
    P(r|L) = \begin{cases}
            \frac{1}{2L^{3}}r^{2}e^{-r/L},\ \text{if}\ r > 0\\
            0,\ \text{otherwise}
        \end{cases}
\end{equation}

and P$(\{\plx\}|r,\{\sigma_{\plx}\})$ is the likelihood of observing the set of parallax measurements given a true cluster distance $r$ and the set of measurement uncertainties and is given by,

\begin{equation}
    P(\{\plx\}|r,\{\sigma_{\plx}\})=\prod_{i=1} \frac{1}{\sqrt{2 \pi \sigma_{\plx_{i}}^{2}}} exp \bigg (- \frac{(\plx_{i} - \frac{1}{r})^{2}}{2 \sigma_{\plx_{i}}^{2}} \bigg ).
    \label{eq:LH}
\end{equation}

We set the length scale `L' within our prior to be 1 kpc and compute the posterior over a distance range of [0.01, 30] kpc.

We sample the posterior distribution and compute the 16$^{th}$, 50$^{th}$, and 84$^{th}$ percentile distances for all 420 open clusters and the open cluster candidates (Section~\ref{sec:serendipitous_ocs}).
Table \ref{tab:cluster_mu_params} contains our distance measurements to the 11 newly confirmed open clusters in Gaia DR2 and our 11 new open cluster candidates. 
The full table of all 420 open clusters mean astrometric parameters and inferred distances will be made available electronically.

We acknowledge that the range between the 16$^{th}$ and 84$^{th}$ distance percentiles may be artificially narrow.
Equation~\ref{eq:LH} implicitly assumes that all cluster members reside at a singular distance rather than occupy a small, realistic range in distance along the line of sight.
The assumption greatly simplifies the calculation and is relatively accurate for distant clusters. However, our treatment may not be the most suitable for very nearby open clusters such as Melotte-25 (the Hyades, \plx $\sim$ 21mas), and Melotte-22 (the Pleiades, \plx $\sim$ 7mas).

\subsection{Verification of our catalog results}\label{ssec:res_validate}

To ascertain some measure of the veracity of our results, we first compare our central cluster parameters and membership lists to previously published membership catalogs.

In addition, we use the more recent Gaia eDR3 data \citep{Brown2021} to determine if the cluster members unique to this work are actual cluster stars or field contamination.

\subsubsection{Comparison with previous results in the literature} \label{sssec:lit_compare}

\begin{figure*}
% \hspace*{-1.5cm}
% \epsscale{1.2}
% \begin{center}
\hspace{-1cm}
\includegraphics[scale=0.75]{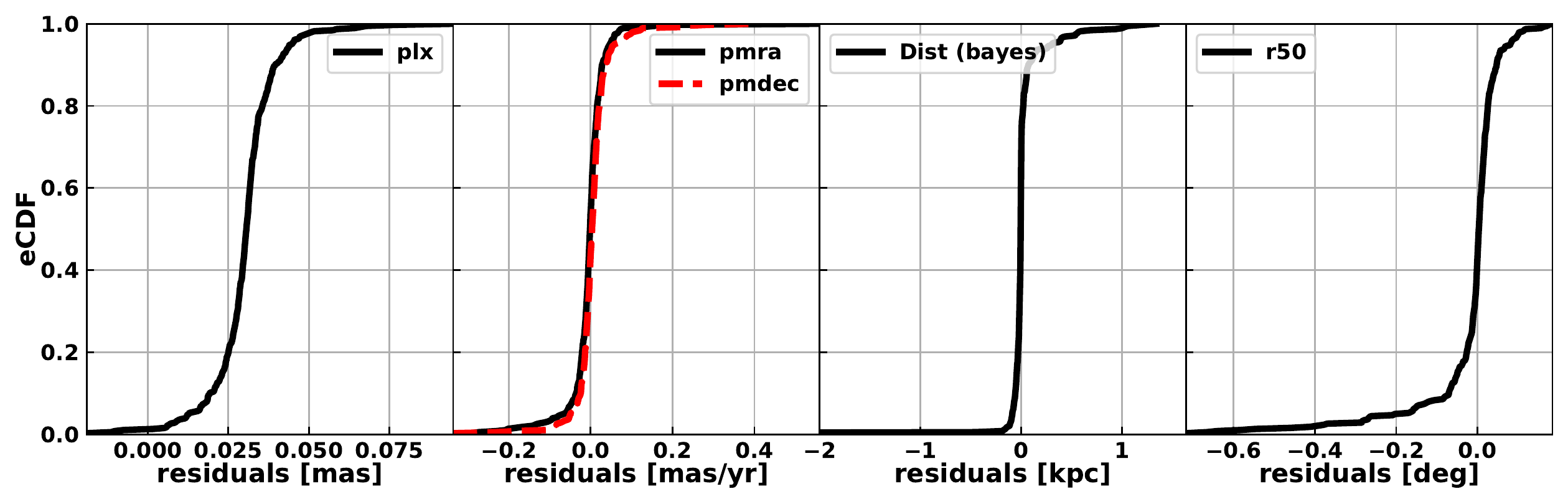}
\parbox{18cm}{\caption{(In order from left to right) Comparison empirical cumulative distribution (eCDF) plots of the mean cluster \plx, \pmra(black solid line) and \pmdec (red dashed line), distance from Bayesian inference (black solid line), and on-sky size containing half of all cluster members (r50). The \plx, \pmra, \pmdec, distance from Bayesian inference, and r50 eCDF plots are comparing the 386 clusters this work has in common with \citet{Cantat-Gaudin2019}. There is overall excellent agreement between mean open cluster parameters in our catalog with the values in the literature from different methods.  \label{fig:dist_one_to_one}   }}

\end{figure*}

We can further scrutinize our results by comparing the mean cluster parameters for our sample with those clusters which have entries within the literature. 
We plot the empirical cumulative distribution functions(eCDF) of the residuals between each cluster's parameter calculated in this work with the literature parameters in Figure \ref{fig:dist_one_to_one}.

The difference in \plx, \pmra, and \pmdec\
was computed for $392$ clusters this work has in common with \citet{Cantat-Gaudin2020}. 
We find that the 
%residuals of
difference in \plx, \pmra, and \pmdec\ is quite small, with median values of 0.031 mas, -0.002 mas yr$^{-1}$, and 0.004 mas yr$^{-1}$, respectively. 
The offset of 0.031 mas in the \plx\ appears to come from our application of the $-0.029\ $ mas zero-point offset (see Section \ref{ssec:data_prep}). Removing this zero-point offset from the cluster \plx\ values leads to a median difference of only $0.002$\ mas with the literature values. This level of agreement with a study using a completely different cluster identification algorithm is strong evidence that we successfully recover the central location of our target clusters.

We find good agreement between our inferred cluster distances and those from \citet{Cantat-Gaudin2019}. The median difference in distance is $-0.003\ $ kpc. 
The range between the 84$^{th}$ and 16$^{th}$ percentiles in the distance discrepancy is only 0.036 kpc.

We also find good agreement in r50, the radius containing half the cluster members. 
%The common with the \citet{Cantat-Gaudin2020}. 
The median difference in r50 between the results of \citet{Cantat-Gaudin2019} and our own is 0.003 deg. We do report much smaller r50 values for about 10\% of clusters (Figure \ref{fig:dist_one_to_one}, bottom left, x \textless -0.10 deg).
Upon investigation, we find that these clusters, according to \citet{Cantat-Gaudin2020}, extend beyond the aperture radius employed in this work to define the input target field (Section \ref{sec:data}).
Thus, the cluster members recovered in this work may populate as wide a range in position for the clusters with large angular sizes. However, based on the agreement in cluster mean \plx, \pmra, and \pmdec, this does not appear to affect our ability to recover the bulk of the clusters in question.

The general agreement in cluster mean parameters suggests that our cluster member lists have significant overlap with those of \citet{Cantat-Gaudin2020}. We compare our catalog with the catalog of \citet{Cantat-Gaudin2020} using Gaia DR2 source ids at a membership probability of $>25\%$, $>50\%$, and $>95\%$. We correct the aperture size to match our target field aperture.

At the lowest level of membership probability we find our that our cluster member list contains 98.7$\%$ of the members reported by \citet{Cantat-Gaudin2020} at the same membership probability according to their method. This overlap decreases to 96.1$\%$ at the next membership level. The overlap decreases to 81.1$\%$ at the highest level of membership.

The overlap in cluster members is a strong argument that our cluster member designation is largely accurate, since our method to construct the cluster membership lists differs from that of \citet{Cantat-Gaudin2020}.
The alternative, that our discrepant methods share the same false positives, appears unlikely.

We find that member stars in our catalog that are not found in \citet{Cantat-Gaudin2020} are fainter than the member stars shared in common between our two catalogs.
The median Gaia DR2 G-band magnitude of common member stars with \pmem$>50\%$ is $15.67$.
The unique member stars found in this work have a median Gaia DR2 G-band magnitude of 16.77, a difference of 1.1 magnitudes. 
At $\pmem>95\%$, the median Gaia G-band of unique member stars identified in this work is still 0.6 magnitudes fainter than the highly probable members common to both catalogs. 
This could indicate that the XDGMM fit is more sensitive to the fainter stars in each target field, given the full treatment of each star's individual covariance matrix.

\subsubsection{Hertzsprung-Russell Diagrams of our Open Clusters}\label{sssec:open_cluster_hrds}
Open clusters populate isochrones within the Hertzsprung-Russell Diagram (HRD). The HRD is a well-known figure that relates a star's evolutionary state to its color and absolute magnitude, in lieu of temperature and luminosity measurements, respectively. We can check the reliability of our method of selecting open cluster groups by placing them on the HRD and checking to see the bulk population properties of our member catalog.

We calculate the absolute magnitudes of each cluster star prior to placing it on the HRD. 
The majority of the open clusters considered in this work reside within the galactic disk of the Milky way, occupying the galactic latitude ({\it{b}}) range [-20,20].
We need to correct for reddening and extinction due to galactic dust, as well as distance, to be able to place the cluster member stars on the HRD accurately. 

Dust extinction and distance has already measured for 392 clusters in our recovered sample of 420 (93.3\%) by \citet{Cantat-Gaudin2020}. In \citet{Cantat-Gaudin2020} a neural network was developed and trained on CMDs of synthetic clusters constructed from 347 real cluster parameters in the literature, the bulk of which came from \citet{Bossini2019a}.

We use the polynomial fit equations of \citet{Babusiaux2018ObservationalDiagrams} to determine the Gaia DR2 passband extinction values, A$_{G}$, A$_{BP}$, A$_{RP}$ for all stars in the 392 clusters with A$_{V}$ measurements from \citet{Cantat-Gaudin2020}. However, some stars have poor photometry such that the Gaia extinction coefficients could not be calculated. 
We circumvent this by selecting a high quality sample of stars from each cluster. We select the stars within each cluster with a probability \textgreater\ 0.85. We also exclude all stars with \textit{gaiadr2.phot\_bp\_rp\_excess\_factor} \textgreater 1.5+0.03$\cdot$(\bp\ - \rp)$^{2}$ as this is indicative of large systematic errors within the \bp\ and \rp\ colors \citep{Riello2018GaiaData}.
The median A$_{G}$, A$_{BP}$, A$_{RP}$ values of these high quality cluster stars is then used to de-extinct and de-redden stars all other stars in each cluster. This approach allows us to apply dust corrections to all probable cluster members, even those with more problematic photometry. However, this means we do not account for differential reddening present in very distant or heavily extincted open clusters.

\begin{figure*}
% \epsscale{1.2}
\hspace{-1.25cm}
\includegraphics[scale=0.75]{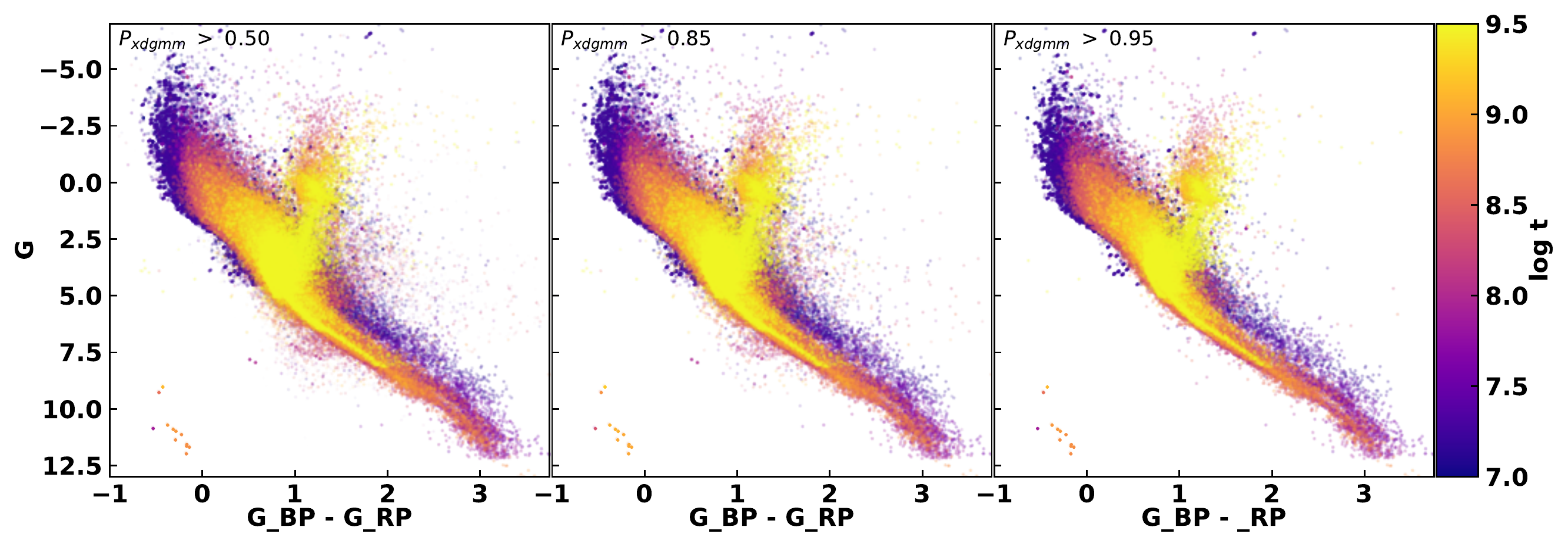}
\parbox{18cm}{\caption{Hertzsprung - Russell diagram of all member stars within the 392 open clusters recovered in this work that also have predicted ages, extinctions, and distances in \citet{Cantat-Gaudin2020}. All stars plotted have a \bp\ - \rp\ excess factor less than 1.5+0.03$\cdot$(\bp\ - \rp\ )$^2$. Stars with at least a  0.25 membership probability are plotted in the left panel, the 0.50 membership level stars are plotted in the middle panel, and the 0.95 membership level The stars are plotted in the right panel. The three Gaia passbands (\g, \bp, \rp) have been corrected for extinction and reddening using the methods described in \citet{Babusiaux2018ObservationalDiagrams}. We find that our use of XDGMM is able to properly capture multiple phases of open cluster evolution with no apparent biases. \label{fig:three_panel_hrd} }}
\end{figure*}

We plot the extinction-corrected absolute \g\ band magnitude as well as the de-reddened \bp\ - \rp\ color of cluster members in Figure \ref{fig:three_panel_hrd}. We plot the HRDs of stars with membership probabilities greater than $50\%$ (left panel), greater than $85\%$(middle panel), and greater than $95\%$ (right panel). We also color code the stars by their parent cluster age as predicted in \citet{Cantat-Gaudin2020}.

Our HRD contains various groups at different stages of stellar evolution, such as the main-sequence, main-sequence turn off, and red-giant branch. All of these features are clearly visible even at the $50\%$ membership probability threshold(Figure \ref{fig:three_panel_hrd}, left panel). 
We also recover 12 white dwarfs as high probability cluster members with this application of XDGMM. 5 are apparent members of the Beehive cluster (NGC-2632), 3 are in the Hyades cluster (Melotte-25) while the Coma star cluster (Melotte-111), the Pleiades (Melotte-22), Graff's cluster (IC-4756), and the Pincushion cluster (NGC-3532) each have 1 white dwarf. 

The overwhelming majority of the cluster member stars appear in physically sensible regions of the HRD.
While it is true that a group of stars sharing the same parallax will form a diagonal sequence within the HRD, our general and close agreement with the mean cluster parameters of previous studies (Section ~\ref{sssec:lit_compare}) is good evidence that we do not claim to find a known cluster at an incorrect astrometric position. 
Instead, the lack of stars at unphysical locations in the diagram, even at \pmem$>50\%$,  suggests that our membership classification does not suffer from significant field contamination due to erroneous or uncertain parallax measurements.
Stars with intrinsic parallax different from the target cluster would appear above or below the main locus. 

The clear age gradient and features in the HRD offer more evidence that our cluster member designations are largely accurate. 
The main sequence turn-off in older clusters in clearly more red and dim than in younger clusters. If we erroneously labeled all stars near the cluster parallax as cluster members, the correlation between cluster age and turn-off position may have been weakened by field interlopers.
In addition, we only find stars along the pre-main-sequence in the youngest clusters.
The age labels of each cluster are taken from \citet{Cantat-Gaudin2020}, who use an entirely different method of determining cluster membership.
The cluster ages shown in Figure~\ref{fig:three_panel_hrd} are completely decoupled from our cluster membership designations, yet the age-dependent features above remain.

Our cluster membership probabilities are calculated independent of any photometry, but we find a correlation between membership probability and position on the HRD relative to the expected locations predicted by cluster isochrones.
A small percentage of stars appear between the main sequence and white dwarfs; others are redder than the red giant branch.
These outliers are far less prominent in the \pmem$>95\%$ panel.
This correlation is clearly apparent for stars just below the main sequence ($\bp\ -\rp\ \approx 1$ and $\g\approx 7.5$; $\bp\ -\rp\ \approx 0$ and $\g \approx 2.5$), colder than the main sequence turnoff, and with intermediate colors relative to the red clump and upper main sequence.

\subsubsection{Membership Testing with Gaia eDR3}\label{ssec:gaia_edr3_check}

Gaia eDR3 \citep{Brown2021} yields an opportunity to validate our cluster membership lists.
The Gaia eDR3 astrometric data is nominally at higher precision than Gaia DR2 and can serve as a check on the quality of our cluster stars memberships.
Specifically, we can statistically test for cluster membership by comparing the Gaia DR2 and eDR3 parallax measurements of the same star.

We first assume that the observed parallax does not suffer from
systematic uncertainty nor bias and is normally distributed about the true
parallax in both Gaia eDR3 and DR2. The observed parallax in either data release is then \citep[as in][]{Luri2018},

\begin{equation}
    p(\plx_j | \plx_{\mathrm{true}}) \propto \normal{\plx_{\mathrm{true}}}{\sigma_{\plx_j}}
    \label{eq:plx}
\end{equation}

where the subscript $j \in$ [DR2, eDR3] refers to the specific data release, $\plx_{\mathrm{true}}$ is the true parallax of the star,
\(\mathcal{N}(\mu, \sigma^2)\) represents a normal distribution with
mean \(\mu\) and variance \(\sigma^2\), and
$\sigma_{\plx_j}$ is the uncertainty of the parallax measurement in each data release .

The relationship between the parallax measurements in both Gaia data sets and the parallax (distance) of the cluster will be different for cluster and field stars. The true distance of a cluster corresponds to a singular cluster parallax, $1/ d_{\mathrm{clus}} = \plx_{\mathrm{clus}}$, if we ignore the small intrinsic width of the cluster. 
The observed parallax of a star relative to the cluster parallax is simply $\delta_{\plx_j} = \plx_j - \plx_{\mathrm{clus}}$, where $j$ again identifies the specific Gaia data release.
For a bonafide cluster member, $\plx_{\mathrm{true}} = \plx_{clus}$ and thus one expects the eDR3 parallax measurement to be closer to the true cluster parallax; \ie, $|\delta_{\plx_{\mathrm{eDR3}}}| < |\delta_{\plx_{\mathrm{DR2}}}|$ if $\sigma_{\plx_{eDR3}} < \sigma_{\plx_{DR2}}$.
However, $\plx_{\mathrm{true}} \neq \plx_{clus}$ for field stars in which case we expect $|\delta_{\plx_{\mathrm{eDR3}}}| \sim |\delta_{\plx_{\mathrm{DR2}}}|$.

The \emph{difference} between Gaia eDR3 and DR2 measurements of the observed parallax relative to the cluster parallax, $\delta_{\plx_{\mathrm{eDR3}}} - \delta_{\plx_{\mathrm{DR2}}}$ should therefore be different for cluster and field stars. 
We define this change in relative parallax $\Delta\plx$ for a single star in terms of the relative Gaia DR2 parallax measurement as,

\begin{equation}
\begin{aligned}
\Delta\plx &= \delta\plx_{\mathrm{eDR3}} - \delta\plx_{\mathrm{DR2}}  \\
 &= \plx_{\mathrm{eDR3}} - \plx_{\mathrm{DR2}}  \\
 &= \normal{0}{1}\sigma_{\plx_\mathrm{eDR3}} - \normal{0}{1}\sigma_{\plx_\mathrm{DR2}}  \\
 &= \normal{0}{1}\sigma_{\plx_\mathrm{eDR3}} - (\delta\plx_{\mathrm{DR2}} - \plx_{\mathrm{true}} + \plx_{\mathrm{clus}}) 
  \label{eq:deltaplx}
\end{aligned}
\end{equation}
where the final two equalities make use of $\normal{\mu}{\sigma^2} \equiv \normal{0}{1}\sigma + \mu$ and our definition of $\delta\plx_{\mathrm{DR2}}$.

When the true parallax of a star is different from $\plx_{\mathrm{clus}}$, as is the case for field stars, $\Delta_\plx$ will have no dependence on $\delta\plx_{\mathrm{DR2}}$ as $\delta\plx_{\mathrm{DR2}} \propto \plx_{\mathrm{true}} - \plx_{\mathrm{clus}}$. 
Cluster stars, however, will show a strong correlation between $\Delta_\plx$ and $\delta\plx_{\mathrm{DR2}}$.
Since $\plx_{\mathrm{true}} = \plx_{\mathrm{clus}}$ for cluster members,   $-\delta\plx_{\mathrm{DR2}} = -\normal{0}{1}\sigma_{\plx_\mathrm{DR2}}$ and thus,  $\Delta_\plx \propto -\delta\plx_{\mathrm{DR2}}$.

\begin{figure}[!hbt]
\hspace*{-.75cm}
% \epsscale{1.2}
\includegraphics[scale=0.58]{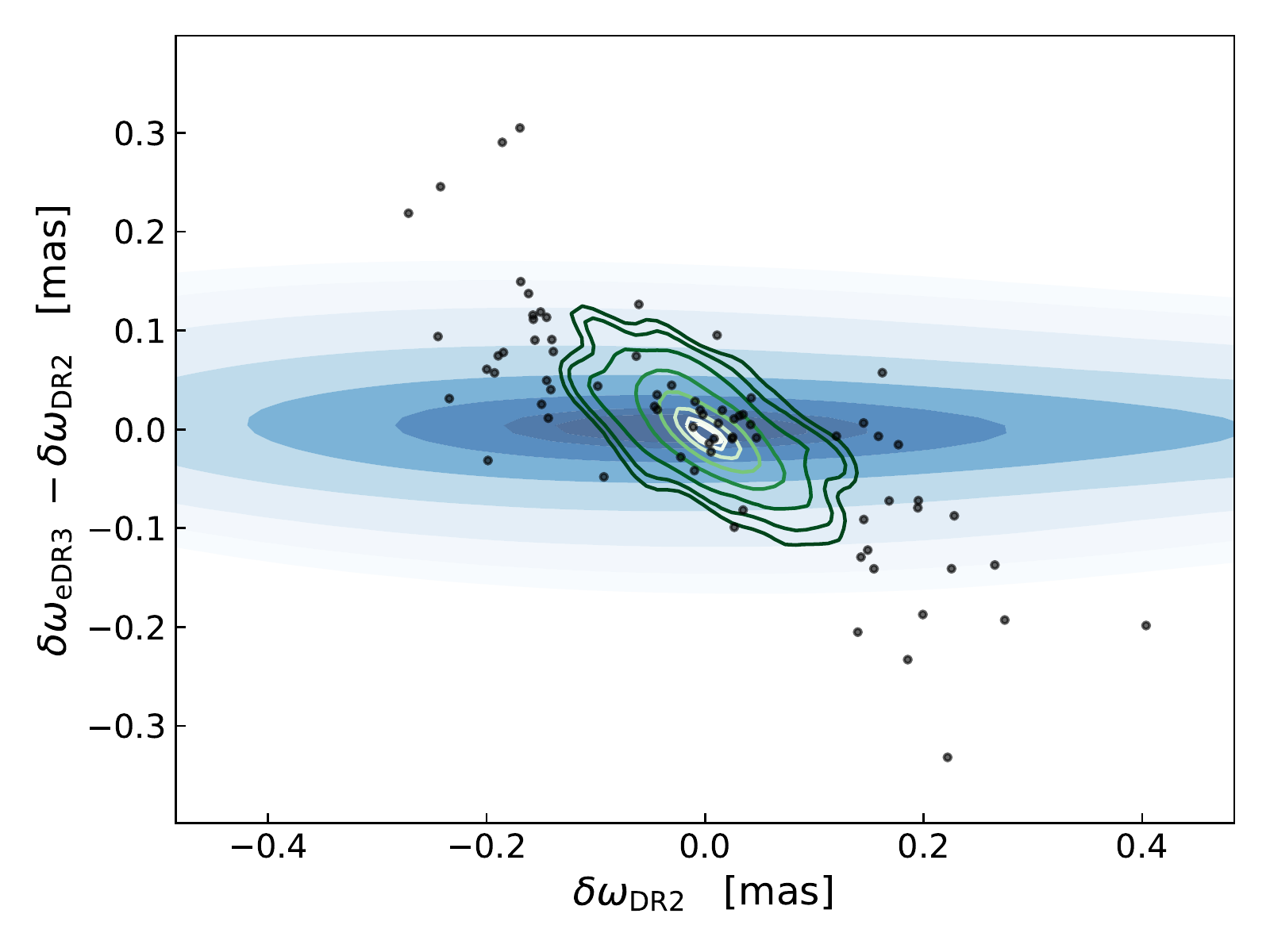}
\parbox{\columnwidth}{\caption{The observed parallax relative to the cluster in Gaia DR2 ($\delta\plx_{\mathrm{DR2}}$) compared to the change in this relative parallax between Gaia eDR3 and DR2 for the target field containing the cluster NGC 188. Field stars (filled contours, darker colors represent higher density) show almost no correlation between these quantities. Bonafide cluster members (open contours, lighter colors denote higher density) show a strong negative correlation. Stars identified as cluster members in our analysis but absent from \citet{Cantat-Gaudin2020} are shown as individual points. See text for details. \label{fig:delta_parallax}  }} 

\end{figure}

Figure~\ref{fig:delta_parallax} shows the distribution of $\Delta\plx$ and $\delta\plx_{\mathrm{DR2}}$ for the field stars and member stars in NGC-188 line-of-sight.
% Here, we examined the stars comprising the sightline home to the cluster NGC 188.
We identified field stars as those with $< 1\%$ membership probability; $\Delta\plx$ has little or no dependence on $\delta\plx_{\mathrm{DR2}}$ for these stars (filled contours).
Bonafide cluster members (open line contours) show a strong, negative correlation between $\Delta\plx$ and $\delta\plx_{\mathrm{DR2}}$.
The \emph{bonafide} cluster member designation applies to those stars with cluster membership $> 50\%$ in both our catalog and that of \citet{Cantat-Gaudin2020}.
Using this definition, we are confident that nothing inherent to our cluster membership analysis is driving the correlation seen in Figure~\ref{fig:delta_parallax}.

We statistically assess the relationship shown in Figure~\ref{fig:delta_parallax} across all target fields using simple linear regression\footnote{as implemented by {\textsc{curve\_fit}}, a part of the {\textsc{scipy}} python package}.
In each target field, we fit a line to $\Delta\plx$ vs.~\ $\delta\plx_{\mathrm{DR2}}$ for three distinction populations -- field stars, bonafide cluster members, and unique cluster members.
Unique cluster members are probable members ($>85\% $ membership probability) according to our analysis but have zero cluster membership probability according to \citet{Cantat-Gaudin2020} (see the black points in Figure~\ref{fig:delta_parallax} for an example).
We attempt to match all stars considered in our XDGMM fits with its Gaia eDR3 counterpart using the crossmatch provided by the Gaia team (\textit{gaiaedr3{.}dr2\_neighbourhood}). 
To ensure a clean match, we demand that the eDR3 target is within $0.15\ \mathrm{mas}$ of the DR2 source on the sky and that the eDR3 and DR2 Gaia magnitudes do not differ by more than $0.1\ \mathrm{mag}$.

The resulting catalog contains $\sim 1.7\times10^6$ stars.
Along each line of sight, we assume the true cluster parallax is the median observed parallax of the bonafide cluster members.
We require at least $20$ bonafide members to ensure a robust measurement of $\plx_{\mathrm{clus}}$ and at least $15$ members of any other group to perform the fitting procedure.
Finally, the intercept of the linear fit must be near zero to guard against significant differences in the observed parallax systematics between the two data releases (see below for details).
These requirements limit our analysis to $307$ sightlines for the bonafide cluster member and field populations and $88$ sightlines for the unique member stars.
The linear fit is performed across the same domain for all three groups. The fit only applies to stellar measurements within the extent of $\delta\plx_{\mathrm{DR2}}$ for cluster stars, including both bonafide and unique members.
Performing the regression across the same domain ensures the most straightforward comparison between cluster and field stars.

\begin{figure}%[!hbt]
%\hspace*{-.75cm}
% \epsscale{1.2}
\includegraphics[scale=0.58]{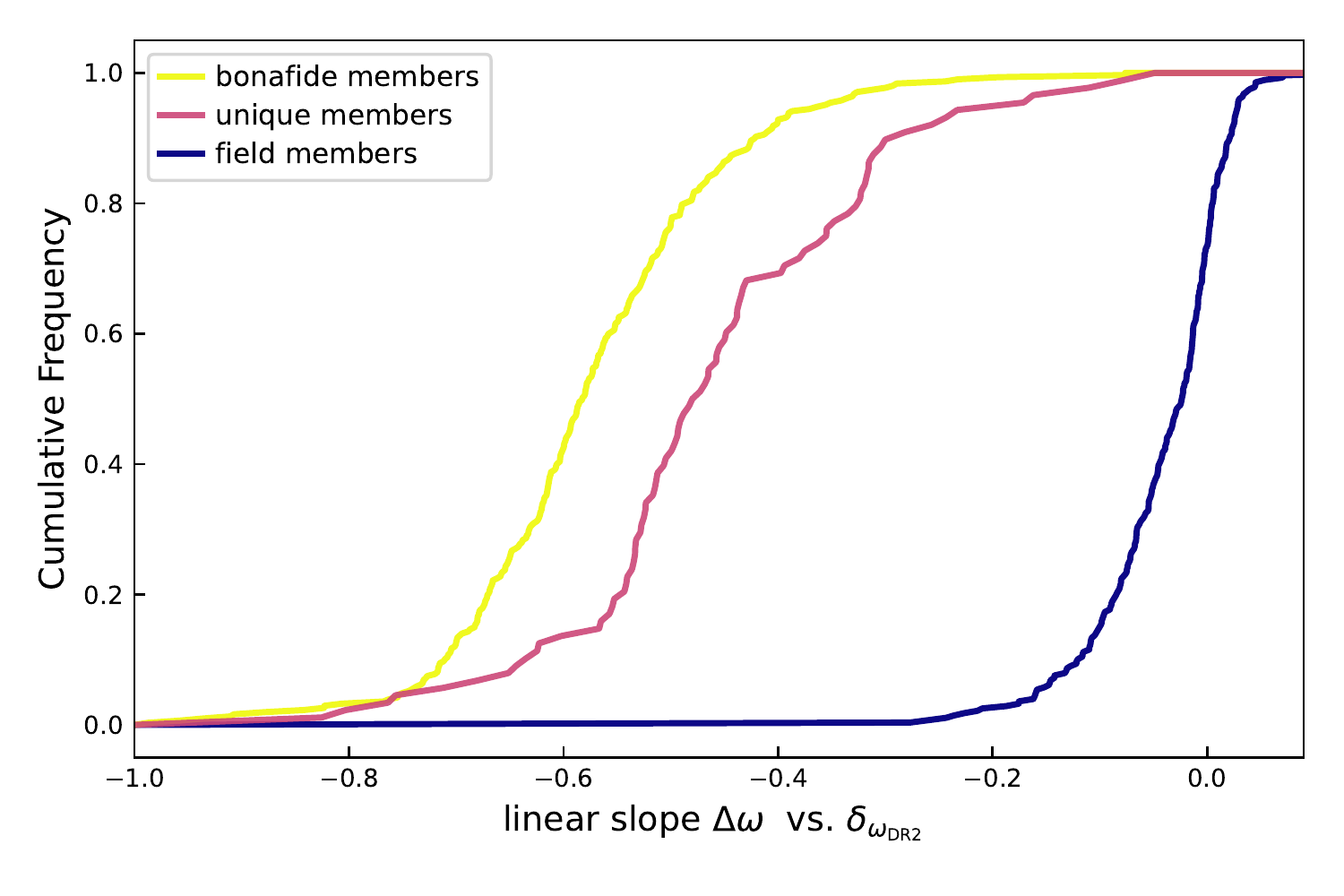}
\parbox{\columnwidth}{\caption{The distribution of linear slope fit to the $\Delta\plx$ vs.~\ $\delta\plx_{\mathrm{DR2}}$ data within three stellar populations along each line of sight. The slope measures the degree of correlation between the two quantities; each line represents the empirical cumulative distribution function of slope within a population. The slope of bonafide cluster members (yellow; median of $-0.58$) is significantly more negative than that of field stars (blue; median of $-0.02$). The distribution of slope measured for cluster members uniquely identified in this work (magenta; median of $-0.48$) is intermediate to the field and bonafide member distributions. The similarity of the bonafide and unique cluster member slope distributions suggests that a majority of the unique cluster members are likely genuine. See text for details. \label{fig:slope} }} 

\end{figure}

We show the distribution of the $\Delta\plx$ vs.~\ $\delta\plx_{\mathrm{DR2}}$ slope measured from the above regression for field, bonafide cluster member, and unique cluster member stars in Figure~\ref{fig:slope}.
The slope of the bonafide cluster members (yellow line) is significantly negative with a median slope of $-0.58$.
The bonafide cluster members show a strong negative correlation between the dependent and independent quantities as predicted by equation~\ref{eq:deltaplx}.
However, the expected slope for a pure cluster member sample is $-1$.
We discuss potential reasons for this deviation below.
Field stars (blue line) show almost no correlation between $\Delta\plx$ and $\delta\plx_{\mathrm{DR2}}$; their median target field slope is $-0.02$.
Given our assumptions, the expected slope value for field stars is zero and near our actual result.
The distribution of slopes measured for unique cluster members (pink line) is much closer to that of bonafide cluster members than that of field stars; the median slope is $-0.48$.

Figure~\ref{fig:slope} presents strong evidence that the majority of unique cluster members genuinely belong to a cluster.
The unique cluster members clearly show a correlation between the observed parallax in Gaia eDR3 and Gaia DR2 that is not seen in the field stars. 
The most negative slope recorded for field stars is $-0.28$ for any line of sight.
The measured slope of the unique cluster member stars is $<-0.28$ in over $90\%$  of all lines of sight considered.
Further, the similarity of the slope distributions of the bonafide and unique cluster member stars suggests that they are largely drawn from the same distribution of cluster-relative parallax ($\delta\plx_{\mathrm{DR3}}$, $\delta\plx_{\mathrm{DR2}}$). 
However, the distribution of slopes for the unique cluster stars is not as negative as that of the bonafide cluster members.
We therefore conservatively conclude between $5\%$ and $45\%$ of the unique cluster stars are possible false positives.
We identify $9{,}118$ unique cluster member stars meet the data criteria detailed within this subsection and more than $7{,}500$ probable unique cluster member stars were removed from this analysis. 
The increase in probable cluster membership due to our unique cluster stars is significant even if the false positive rate is near the high-end of our estimated range.

To simplify the calculation of $\Delta\plx$ and related quantities, we made several assumptions that are likely incorrect in detail but do not detract from our line of reasoning. 
The Gaia reported parallax contains some systematic error and is likely biased. 
In our calculations, we do correct for the global parallax offset of $0.029 \mathrm{mas}$ in Gaia DR2 \citep{lindegren2018astrometric} and $0.017 \mathrm{mas}$ in Gaia eDR3 \citep{Lindegren2020GaiaPosition} but then assume no additional systematic errors.
A significantly non-zero intercept of the linear fit to $\Delta\plx$ vs.~\ $\delta\plx_{\mathrm{DR2}}$ along a sightline suggest influential and dissimilar bias in the parallax measurements of the two Gaia data releases.
To mitigate this where possible, we remove any sightline in which the absolute value of the intercept is greater than one-tenth the median uncertainty in the observed parallax of the population under consideration.
All of the reported statistics from Figure~\ref{fig:slope} include this restriction.
Finally, clusters do have intrinsic width, often near the scale of $\delta_{\plx}$ for cluster members.
The combination of these effects contributes to the change in the slope distribution for bonafide and unique cluster members relative to the expectation of $-1$ and likely contributes to the slight negative skew seen in Figure~\ref{fig:slope} for field stars.
Still, the slope distributions of the bonafide member and field stars are sufficiently different from that of field stars to support our conclusions above.

\section{Previously unpublished clusters discovered with XDGMM}\label{sec:serendipitous_ocs}
\par 

We report the serendipitous discovery of 11 previously unpublished open cluster candidates.
These candidates were found as components with relatively low differential entropy within the best-fit XDGMM of the 420 target fields with a successful recovery of the central target cluster. When looking for candidates in a given field, we considered all non-target cluster components of the best-fit model that had a differential entropy within a factor of 4 of the target cluster. We find 97 components that meet this differential entropy requirement.

The low differential entropy components must pass several tests before we consider them cluster candidates. First, we check to see that the cluster is spatially compact as well compact in proper-motion. 

We check compactness in proper-motion by employing the proper-motion dispersion check developed by \citet[][section 5.3]{Cantat-Gaudin2019}. We calculate the total dispersion in proper-motion for each component and compare it to their analytical threshold. We examine the on-sky spatial compactness of each candidate using the method of \citet[][section 2.3]{Cantat-Gaudin2018a}. We compare the sum of the minimum spanning tree branch lengths of the candidate cluster member positions to those of a randomly generated position distribution of the same size. We remove 44 candidates that do not meet both these requirements.

We then examine the parallax and sky position distributions of the remaining 53 candidates.
We calculate the slope of $\Delta_{\plx}$ vs.~\ $\delta_{\plx_{\mathrm{DR2}}}$ using the probable members of each component (see Section~\ref{ssec:gaia_edr3_check} for details).
Only 3 candidates do not have a slope $<-0.4$, which is more negative than the field members within any target field (Figure~\ref{fig:slope}).

We then determine if any of the 50 candidates were previously reported using Gaia DR2 data.
To do so, we checked for overlap between the Gaia DR2 source IDs of our candidate members and the source IDs of membership lists from the literature.
We compiled a list of all unique source IDs reported as members of \emph{any} cluster by \citet{Cantat-Gaudin2018b}, \citet{Cantat-Gaudin2019GaiaPerseus}, \citet{Castro-Ginard2018}, \citet{Castro-Ginard2019} \citet{Castro-Ginard2020HuntingDisc},  \citet{Ferreira2020DiscoveryDR2}, \citet{Sim2019}, \citet{Liu2019}, or \citet{Hunt2020ImprovingData}. 
One or more of these studies previously identified a cluster with members from 39 of our candidates. Only 11 previously unreported candidate clusters remained.

Finally, we estimate the physical size of each candidate.
We calculate a physical r50 value, the radius encompassing $50\%$ of the candidate member stars, using the candidate's bayesian inversion parallax distance (see \ref{ssec:cluster_param}).
The 11 remaining candidates span a physical r50 value ranging from 1.15 parsecs to 11.3 parsecs, placing them in the bulk of typical open cluster sizes \citep[see][Figure 4]{Cantat-Gaudin2019}.

The final 11 newly discovered open cluster candidates all have different parent target fields.
The gaussian component associated with the candidate had either the lowest (6 of 11 fields) or second-lowest differential entropy of the best-fit components.

We stress that these 11 systems are open cluster \emph{candidates}. The member stars of these systems occupy a relatively small, compact volume in the space of \plx, \pmra, and \pmdec.
All candidates passed the statistical tests mentioned above and their qualitative characteristics (see below) are consistent with being open clusters.
Still, confirming these candidates as bonafide open clusters will require dedicated observational follow-up and detailed investigation. The 11 clusters are called XDOCC (eXtreme Deconvolution Open Cluster Candidates) and numbered [1-11].

We present the color-magnitude diagrams of the 11 new open cluster candidates in Figure \ref{fig:new11_xdocc_cluster_candidates_cmds} and report their central parameters in Table~\ref{tab:cluster_mu_params}. We get age, distance, and extinction estimates for the 11 cluster candidates by using the Auriga Neural Net developed by \citet{Kounkel2020}. The Auriga neural net has been trained on Gaia DR2 photometry [\g, \bp, \rp], 2MASS \citep{Skrutskie2006The2MASS} photometry [J, H, K], and Gaia DR2 parallaxes in order to predict cluster age, distance, and interstellar extinction (A$_V$). We also plot isochrones using the derived parameters from Auriga, assuming solar metallicity in Figure~\ref{fig:new11_xdocc_cluster_candidates_cmds}. We describe the 11 new candidates and their predicted Age, A$_{V}$, and distance from the Auriga Neural Net below.

\subsubsection{XDOCC-01}\label{ssec:XDOCC_01}
XDOCC-01 (Figure \ref{fig:new_xdocc_clusters_page1}, row 1) is in the target field of NGC-457. Its proper-motion overlaps with that of NGC-457. The cluster is distinct from NGC-457 within parallax-magnitude space. XDOCC-01 has an age of $\log t=8.53\pm$ 0.08, an extinction value of A$_{V}$=0.70 $\pm$ 0.09, and a distance of 813.4 $\pm$ 23.8pc.

\subsubsection{XDOCC-02}\label{ssec:XDOCC_02}
XDOCC-02 (Figure \ref{fig:new_xdocc_clusters_page1}, row 2) is located within the target field of NGC-1960. It has a clear distinct grouping within proper-motion space from NGC-1960. XDOCC-02 does overlap with NGC-1960 within parallax-magnitude space. XDOCC-02 has an age of $\log t=7.54\pm$ 0.09, an extinction value of A$_{V}$=0.57 $\pm$ 0.05, and a distance of 1330.6  $\pm$ 53.2pc.

\subsubsection{XDOCC-03}\label{ssec:XDOCC_03}
XDOCC-03 (Figure \ref{fig:new_xdocc_clusters_page1}, row 3) is in the LOS of NGC-2345. It is closer than NGC-2345 in parallax, with a clear separation in parallax-magnitude space. It is also a distinct group in proper-motion. XDOCC-03 has an age of $\log t=7.28\pm$ 0.09, an extinction value of A$_{V}$=0.71 $\pm$ 0.07, and a distance of 1175.8 $\pm$ 56.0pc.

\subsubsection{XDOCC-04}\label{ssec:XDOCC_04}
XDOCC-04 (Figure \ref{fig:new_xdocc_clusters_page1}, row 4) is located within the target field of NGC-2482. Its proper-motion grouping is separate and distinct from that of NGC-2482. The parallax-magnitude distribution is also separate from the distribution of NGC-2482, appearing to be farther away. XDOCC-04 appears to be farther away than NGC-2482. XDOCC-04 has an age of $\log t=7.78\pm$ 0.12, an extinction value of A$_{V}$=0.53 $\pm$ 0.12, and a distance of 2891.2 $\pm$ 107.6pc.

\subsubsection{XDOCC-05}\label{ssec:XDOCC_05}
XDOCC-05 (Figure \ref{fig:new_xdocc_clusters_page2}, row 1) is in the target field of NGC-2533. It has a distinct separation in proper-motion. It has a slight separation from NGC-2533 in parallax-magnitude space, appearing to be closer. XDOCC-05 has an age of $\log t=8.18\pm$0.12, an extinction value of A$_{V}$=0.24 $\pm$ 0.05, and a distance of 1575.8 $\pm$ 59.7pc.

\subsubsection{XDOCC-06}\label{ssec:XDOCC_06}
XDOCC-06 (Figure \ref{fig:new_xdocc_clusters_page2}, row 2) is located within the target field of IC-2395. It has a distinct grouping in proper-motion space, as well as in parallax-magnitude space. XDOCC-06 appears to be farther away than IC-2395. XDOCC-06 has an age of $\log t=7.93\pm$ 0.18, an extinction value of A$_{V}$=1.02 $\pm$ 0.12, and a distance of 1399.1 $\pm$ 64.0pc.

\subsubsection{XDOCC-07}\label{ssec:XDOCC_07}
XDOCC-07 (Figure \ref{fig:new_xdocc_clusters_page2}, row 3) is located within the target field of NGC-3324. It has a distinct grouping in proper-motion space, as well as in parallax-magnitude space. XDOCC-07 appears to be closer than NGC-3324. XDOCC-07 has an age of $\log t=7.86\pm$ 0.08, an extinction value of A$_{V}$=0.58 $\pm$ 0.05, and a distance of 460.1 $\pm$ 10.0pc.

\subsection{XDOCC-08}\label{spec:XDOCC_08}
XDOCC-08 (Figure \ref{fig:new_xdocc_clusters_page2}, row 4) is located within the target field of NGC-6514. XDOCC-08 is separate from NGC-6514 in proper-motion space. The parallax-magnitude distribution slightly overlaps with that of NGC-6514, but appears to be slightly farther away. XDOCC-08 has an age of $\log t=7.52\pm$ 0.19, an extinction value of A$_{V}$=2.50$\pm$ 0.23, and a distance of 1511.9 $\pm$ 137.9pc.

\subsubsection{XDOCC-09}\label{ssec:XDOCC_09}
Open cluster XDOCC-09 (Figure \ref{fig:new_xdocc_clusters_page3}, row 1) is located in the Stock-1 target field. It has a distinct grouping in proper-motion, as well as in parallax-magnitude space. XDOCC-09 appears to be farther away in parallax than Stock-1. XDOCC-09 has an age of $\log t =8.00\pm0.09$, an extinction value of A$_{V}$=0.66 $\pm$ 0.08, and a distance of 552.7 $\pm15.6$ pc.

\subsubsection{XDOCC-10}\label{ssec:XDOCC_10}
Open cluster XDOCC-10 (Figure \ref{fig:new_xdocc_clusters_page3}, row 2) is located in the Berkeley-87 LOS. Its distribution in parallax-magnitude space also overlaps with that of Berkeley-87. The only indication of it being a distinct cluster of stars is in proper-motion space, where there is a clear separation between the two groups. XDOCC-10 has an age of $\log t=7.63 \pm$ 0.32, an extinction value of A$_{V}$=3.97 $\pm$ 0.17, and a distance of 1994.9 $\pm$ 144.7pc.

\subsection{XDOCC-11}\label{ssec:XDOCC_11}
Open cluster XDOCC-11 (Figure \ref{fig:new_xdocc_clusters_page3}, row 3) is located in the target field of IC-1369. It is distinct from IC-1369 in proper-motion space. Its parallax-magnitude space distribution very slightly overlaps with IC-1369, but is distinct and closer in distance. XDOCC-11 has an age of $\log t=8.13\pm$ 0.11, an extinction value of A$_{V}$=1.37 $\pm$ 0.10, and a distance of 1761.9 $\pm$ 89.3pc.

\section{Summary} \label{sec:summary}
In this work we present an automated application of extreme deconvolution gaussian mixture models (XDGMMs) to produce membership lists for open clusters. While extreme deconvolution has been used before in open cluster membership determination, it had not been applied within an automated pipeline to a large number (N \textgreater\ 400) of open clusters.

We fit a XDGMM to the Gaia [\plx, \pmra,  \pmdec] measurements within each target field.
Extreme deconvolution is able to account for all the measurement uncertainties, covariances, and correlations within Gaia DR2 astrometric data and predict their intrinsic distribution. A gaussian mixture model is then fit to the underlying distribution.
We successfully identify an open cluster separate from the field within the 3-dimensional astrometric space of \pmra, \pmdec, and \plx\ in 420 target fields out of a total initial sample of 426.  

Comparison with previous literature results suggests our method is successful in identifying cluster members (Section~\ref{sssec:lit_compare}).
We find excellent agreement between mean cluster parameters for the 392 open clusters in our parent sample that are also reported in the catalog of \citet{Cantat-Gaudin2020}. 
In addition, our cluster membership lists includes $98.7\%$ of the previously reported cluster members by \citet{Cantat-Gaudin2020} for the same 392 target clusters.
The combination of extreme deconvolution and gaussian mixture models also found a substantial population of new cluster members.
The new members stars are fainter (an average of $1.1$ magnitudes in $G$) than those in the literature, owing to the use of the individual covariance matrices in model fitting.

Several types of cluster identification are noteworthy.
This work contains the first Gaia DR2-based membership lists for 11 previously known open clusters.
In addition, we fit the entire astrometric volume surrounding each target cluster and identify open cluster candidate systems attributed to gaussian components distinct from the target cluster, finding 50.
Upon review, we find 39 of these candidates have been previously reported as open clusters in the literature.
However, we also report the serendipitous discovery 11 new open clusters candidates within this work (Section~\ref{sec:serendipitous_ocs}).

This work address several properties, drawbacks, or concerns previously associated with using Gaussian Mixture Models (GMMs) to detect open clusters.
The cluster detection performance of GMMs is sensitive to the search volume and the number of model components \citep{Hunt2020ImprovingData}. 
Efficiently applying GMMs or XDGMMs to a large sample of clusters requires not only non-interactive solutions to these concerns but also raises a new problem.
Since the field will be represented by one or more components, any pipeline-like analysis must be able to automatically select the model component associated with the cluster.

To define the initial cluster search volume, we only require an initial central cluster position in the sky, an estimate of the projected cluster size, and any estimates of distances to extract the relevant Gaia DR2 data and perform data preprocessing. Our automated preprocessing routine conservatively defines a sufficient volume in parallax and proper-motion space, centered on the sky coordinates and distance estimates of the target cluster, designed to encapsulate all potential cluster members. This data preprocessing requires little to no fine tuning for any one cluster (Section~\ref{ssec:data_prep}). 

We determine the number of model components and cluster component using quantitative methods.
We select the best-fit XDGMM of each target field by minimizing the BIC; the best-fit model contained a component associated with the target cluster in 414/426 cases.
The model component with the smallest differential entropy, a measure of compactness, was automatically selected as the target cluster.
The minimal $h$ component represented either the target cluster ($385$ target fields), an adjacent open cluster ($22$ target fields), or a new open cluster candidate ($6$ targets fields) in the vast majority of best-fitting models.

The differential entropy is a useful metric in finding and selecting open cluster gaussian mixture components from within a sight line. The agnostic nature of using the differential entropy over an arbitrarily chosen sigma significance will allow for greater flexibility in differentiating the astrometric over-density that constitute most open clusters in the Milky Way.

In this work we have shown how XDGMM can be an effective and capable method to calculate comprehensive membership probabilities across a multitude of open cluster morphologies. Moving forward, we plan to use what we learned from automating fitting with XDGMM to calculate membership lists for a larger sample of known open clusters in the Milky Way in order to fully take advantage of the greater completeness and exquisite data quality that Gaia eDR3 offers.

\section{Acknowledgements}
The authors thank the referee for the comments and suggestions, which have greatly improved the quality of this paper. KOJ thanks the LSSTC Data Science Fellowship Program, which is funded by LSSTC, NSF Cybertraining Grant \#1829740, the Brinson Foundation, and the Moore Foundation; their participation in the program has benefited this work. JB was supported in part by Vanderbilt University through the Stevenson Postdoctoral Fellowship. KHB was supported by NSF Grant AST-1358862.

This work has made use of data from the European Space Agency (ESA) mission {\it Gaia} (\url{https://www.cosmos.esa.int/gaia}), processed by the {\it Gaia} Data Processing and Analysis Consortium (DPAC, \url{https://www.cosmos.esa.int/web/gaia/dpac/consortium}). Funding for the DPAC has been provided by national institutions, in particular the institutions participating in the {\it Gaia} Multilateral Agreement. 

% This work makes use of multiple publicly available python modules, including Astropy, Astroquery, Numpy, Matplotlib, Pandas, Scipy, and Scikitlearn.

This research made use of Astropy, a community-developed core Python package for Astronomy \citep{Price-Whelan2018ThePackage, Robitaille2013Astropy:Astronomy}, Astroquery \citep{Ginsburg2019Astroquery:Python}, the IPython package \citep{Perez2007IPython:Computing}, matplotlib, a Python library for publication quality graphics \citep{Hunter2007Matplotlib:Environment}, NumPy \citep{Harris2020ArrayNumPy}, pandas \citep{McKinney2011Pandas:Statistics, McKinney2010DataPython}, Scikitlearn \citep{Pedregosa2015Scikit-learn}, and SciPy \citep{Virtanen2020SciPyPython}.

% \vspace*{1.5cm}

\bibliography{references}

\floattable
\begin{deluxetable}{lccccccccccccccc}[ht!] 
\tabletypesize{\small}
\tablecaption{Cluster parameters for 11 open clusters confirmed in this work and 11 newly discovered open clusters. All $\sigma$ values in this table are median absolute deviations (MAD) multiplied by the factor 1.4826 to approximate a standard deviation. The 'dmode$_{16}$' and 'dmode$_{84}$' values are the $16^{th}$ and $84^{th}$ percentiles of the distance posterior distribution. 'dmode' is the $50^{th}$ percentile of the distance posterior distribution. The full table of 431 cluster parameters will be made available electronically. \label{tab:cluster_mu_params} }
% \tablewidth{0pt}
\tablehead{
\colhead{Name}
&\colhead{$\alpha$}
&\colhead{$\delta$}
&\colhead{$\varpi$}
&\colhead{$\sigma_{\varpi}$}
&\colhead{$\mu_{\alpha*}$}
&\colhead{$\sigma_{\mu_{\alpha*}}$}
&\colhead{$\mu_{\delta}$}
&\colhead{$\sigma_{\mu_{\delta}}$}
&\colhead{$N_{p > 0.85}$}
&\colhead{$r_{50}$}
&\colhead{dmode$_{16}$}
&\colhead{dmode}
&\colhead{dmode$_{84}$}
&\colhead{R$_{GC}$}
&\colhead{Z$_{GC}$}\\
\colhead{}
&\colhead{deg}
&\colhead{deg}
&\colhead{mas}
&\colhead{mas}
&\colhead{mas/yr}
&\colhead{mas/yr}
&\colhead{mas/yr}
&\colhead{mas/yr}
&\colhead{}
&\colhead{deg}
&\colhead{kpc}
&\colhead{kpc}
&\colhead{kpc}
&\colhead{kpc}
&\colhead{kpc}
}
\startdata
Berkeley-42  & 286.31 &   1.90 & 0.09 & 0.11 & -2.84 & 0.26 & -5.95 &     0.25 &  236 &   0.02 & 10.460 & 11.041 &   11.621 &  6.54 & -0.43 \\
Bochum-1     &  96.33 &  19.90 & 0.23 & 0.06 & -0.19 & 0.08 & -0.43 &     0.07 &   29 &   0.13 &  4.477 &  4.586 &    4.695 & 12.81 &  0.32 \\
Bochum-14    & 270.52 & -23.70 & 0.31 & 0.09 &  0.25 & 0.13 & -1.06 &     0.10 &  107 &   0.13 &  3.050 &  3.083 &    3.116 &  5.25 & -0.01 \\
Bochum-7     & 131.10 & -45.97 & 0.18 & 0.03 & -3.09 & 0.10 &  3.65 &     0.15 &   59 &   0.19 &  5.405 &  5.484 &    5.563 & 10.32 & -0.16 \\
Collinder-96 &  97.62 &   2.84 & 0.12 & 0.07 &  0.13 & 0.29 &  0.38 &     0.36 &   57 &   0.11 &  7.895 &  8.257 &    8.619 & 16.05 & -0.43 \\
NGC-1931     &  82.80 &  34.24 & 0.45 & 0.05 &  0.37 & 0.12 & -1.91 &     0.19 &   27 &   0.09 &  2.227 &  2.262 &    2.298 & 10.55 &  0.04 \\
NGC-2467     & 118.18 & -26.37 & 0.19 & 0.03 & -2.55 & 0.08 &  2.58 &     0.11 &  154 &   0.09 &  5.093 &  5.169 &    5.245 & 11.59 &  0.08 \\
NGC-3247     & 156.71 & -57.93 & 0.32 & 0.11 & -6.49 & 1.20 &  3.02 &     0.60 &  225 &   0.05 &  2.967 &  2.986 &    3.006 &  8.08 &  0.01 \\
NGC-6514     & 270.65 & -22.90 & 0.85 & 0.04 &  0.33 & 0.32 & -1.68 &     0.20 &  113 &   0.23 &  1.169 &  1.174 &    1.179 &  7.14 &  0.02 \\
Trumpler-24  & 254.01 & -40.47 & 0.84 & 0.02 &  0.49 & 0.62 & -1.80 &     0.54 &   19 &   0.49 &  1.194 &  1.202 &    1.209 &  7.15 &  0.06 \\
Trumpler-27  & 264.08 & -33.49 & 0.49 & 0.01 & -0.15 & 0.08 & -1.28 &     0.06 &   20 &   0.20 &  2.003 &  2.032 &    2.061 &  6.28 & -0.01 \\
\hline
XDOCC-01 &  19.90 &  58.35 & 1.26 &   0.06 &  -1.68 &    0.11 &  -0.34 &     0.06 &   22 & 0.17 &   0.788 &  0.796 &    0.805 &  8.80 & -0.03 \\
XDOCC-02 &  84.05 &  34.30 & 0.80 &   0.05 &   0.28 &    0.11 &  -2.29 &     0.07 &   26 & 0.14 &   1.215 &  1.261 &    1.311 &  9.57 &  0.06 \\
XDOCC-03 & 107.01 & -13.05 & 0.88 &   0.04 &  -3.32 &    0.36 &   0.55 &     0.13 &   25 & 0.21 &   1.121 &  1.140 &    1.159 &  8.95 & -0.02 \\
XDOCC-04 & 118.83 & -24.17 & 0.33 &   0.04 &  -2.71 &    0.08 &   3.06 &     0.09 &   60 & 0.21 &   2.848 &  3.023 &    3.221 & 10.06 &  0.14 \\
XDOCC-05 & 121.75 & -29.90 & 0.60 &   0.02 &  -3.66 &    0.05 &   3.56 &     0.03 &   19 & 0.12 &   1.590 &  1.636 &    1.685 &  9.06 &  0.07 \\
XDOCC-06 & 130.93 & -48.15 & 0.77 &   0.04 &  -6.38 &    0.11 &   3.61 &     0.09 &   33 & 0.06 &   1.271 &  1.299 &    1.327 &  8.47 & -0.05 \\
XDOCC-07 & 159.48 & -58.72 & 2.28 &   0.04 & -14.32 &    0.23 &   0.93 &     0.23 &   21 & 0.15 &   0.433 &  0.436 &    0.439 &  8.01 &  0.02 \\
XDOCC-08 & 270.76 & -22.73 & 0.73 &   0.05 &   0.67 &    0.11 &  -2.57 &     0.08 &   51 & 0.12 &   1.320 &  1.348 &    1.377 &  6.79 &  0.01 \\
XDOCC-09 & 293.60 &  25.00 & 1.83 &   0.04 &  -0.46 &    0.10 &  -8.74 &     0.37 &   47 & 0.66 &   0.546 &  0.550 &    0.553 &  8.04 &  0.05 \\
XDOCC-10 & 305.42 &  37.29 & 0.51 &   0.06 &  -2.39 &    0.11 &  -5.23 &     0.12 &   40 & 0.16 &   1.884 &  1.919 &    1.956 &  7.87 &  0.03 \\
XDOCC-11 & 317.73 &  47.72 & 0.47 &   0.04 &  -1.01 &    0.05 &  -1.74 &     0.05 &   21 & 0.10 &   1.999 &  2.063 &    2.131 &  8.54 &  0.02 \\
\enddata
\end{deluxetable}

\floattable
\begin{deluxetable}{lcccccccccccc}[ht!] 
\tabletypesize{\small}
\tablecaption{Individual membership measurements for 10 member stars within 10 clusters within this catalog. All columns except for $Cluster$ and $P_{mem}$ are directly downloaded from the Gaia DR2 archive. The full member star catalog will be made available electronically. \label{tab:membs_params} }
% \tablewidth{0pt}
\tablehead{
\colhead{Cluster}
&\colhead{ID$_{GDR2}$}
&\colhead{$\alpha$}
&\colhead{$\delta$}
&\colhead{$\varpi$}
&\colhead{$\sigma_{\varpi}$}
&\colhead{$\mu_{\alpha*}$}
&\colhead{$\sigma_{\mu_{\alpha*}}$}
&\colhead{$\mu_{\delta}$}
&\colhead{$\sigma_{\mu_{\delta}}$}
&\colhead{$G$}
&\colhead{$G_{BP}-G_{RP}$}
&\colhead{$P_{mem}$}\\
\colhead{}
&\colhead{}
&\colhead{deg}
&\colhead{deg}
&\colhead{mas}
&\colhead{mas}
&\colhead{mas/yr}
&\colhead{mas/yr}
&\colhead{mas/yr}
&\colhead{mas/yr}
&\colhead{mag}
&\colhead{mag}
&\colhead{}
}
\startdata
 Berkeley\_104 &   431611438369749376 &     0.817 &   63.580 &    0.212 &      0.025 &    -2.397 &       0.042 &      0.022 &        0.039 &    14.810 &     1.648 &  1.00 \\
  Berkeley\_66 &   461225237865057792 &    46.072 &   58.710 &    0.163 &      0.072 &    -0.802 &       0.154 &      0.153 &        0.127 &    16.808 &     1.717 &  0.91 \\
  Berkeley\_70 &   194973343236220544 &    81.454 &   41.925 &    0.015 &      0.123 &     0.939 &       0.224 &     -1.927 &        0.160 &    17.469 &     1.362 &  0.90 \\
Collinder\_394 &  4085955053336931840 &   283.095 &  -19.961 &    1.205 &      0.189 &    -1.067 &       0.332 &     -5.679 &        0.273 &    17.821 &     1.780 &  0.74 \\
       IC\_166 &   511352282214681216 &    27.643 &   61.804 &    0.167 &      0.088 &    -1.611 &       0.092 &      1.303 &        0.116 &    17.294 &     1.569 &  0.98 \\
      NGC\_188 &   573748841233408384 &    14.078 &   85.085 &    0.528 &      0.075 &    -2.601 &       0.176 &     -0.775 &        0.110 &    17.243 &     1.242 &  0.97 \\
     NGC\_2099 &  3451182079777549568 &    88.032 &   32.572 &    0.547 &      0.037 &     2.108 &       0.079 &     -5.848 &        0.066 &    10.673 &     0.243 &  0.76 \\
     NGC\_2158 &  3426177016401394944 &    91.734 &   24.098 &    0.113 &      0.164 &    -0.283 &       0.255 &     -2.350 &        0.218 &    17.830 &     1.290 &  0.74 \\
     NGC\_2439 &  5598104289661179648 &   115.143 &  -31.693 &    0.254 &      0.118 &    -2.240 &       0.176 &      3.170 &        0.197 &    17.724 &     1.149 &  0.88 \\
     NGC\_2539 &  5726703685031241344 &   122.592 &  -13.175 &    0.764 &      0.032 &    -2.378 &       0.047 &     -0.593 &        0.032 &    15.039 &     0.804 &  1.00 \\
\enddata
\end{deluxetable}

\newpage
\appendix

\section{Color magnitude diagrams of 11 newly discovered open cluster}
\begin{figure*}
% \epsscale{1.2}
% \begin{center}
\hspace{-1.0cm}
% \vspace{1cm}
\includegraphics[scale=1.]{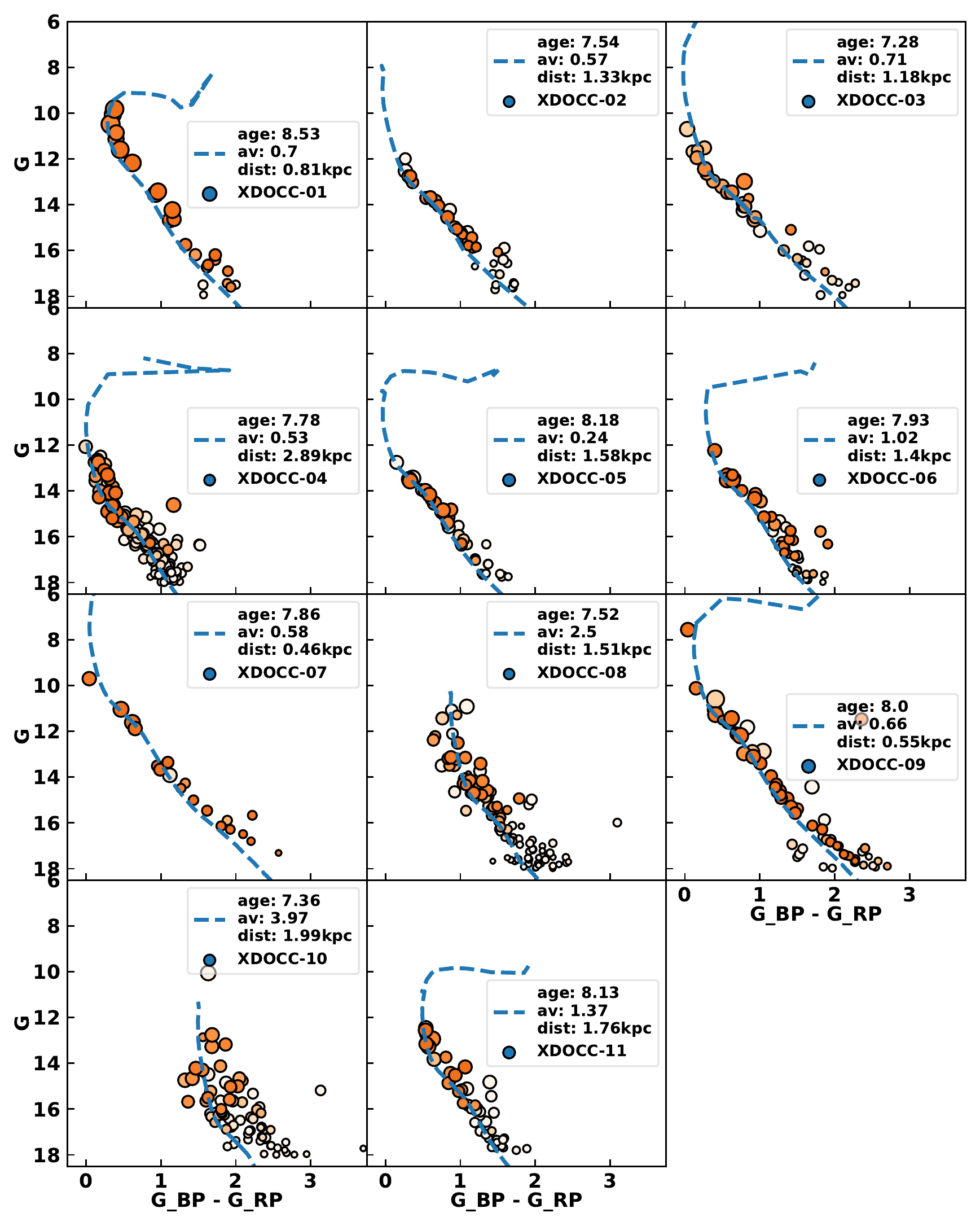}
\parbox{18cm}{\caption{Multi plot color-magnitude diagrams of the 11 newly discovered open clusters within this work. Cluster member stars are plotted as orange colored points. The points are colored by their membership probability, where darker colors == higher probability. The points are sized by their \bp\ - \rp\ uncertainty, where bigger == lower uncertainty. The isochrone age, distance, and extinction values are predictions from the Auriga Neural Net developed by \citet{Kounkel2020}. All isochrones have been set to a solar metallicity. \label{fig:new11_xdocc_cluster_candidates_cmds} }}

% Plotted for each cluster is the absolute G-band magnitude and the dereddened BP-RP color(see Sec \ref{ssec:open_cluster_hrds}).
% \end{center}
\end{figure*}

\section{Four plots of 11 newly discovered open clusters}

\begin{figure*}
% \epsscale{1.2}
% \begin{center}
\hspace{-0.75cm}
% \vspace{1cm}
\includegraphics[scale=0.99]{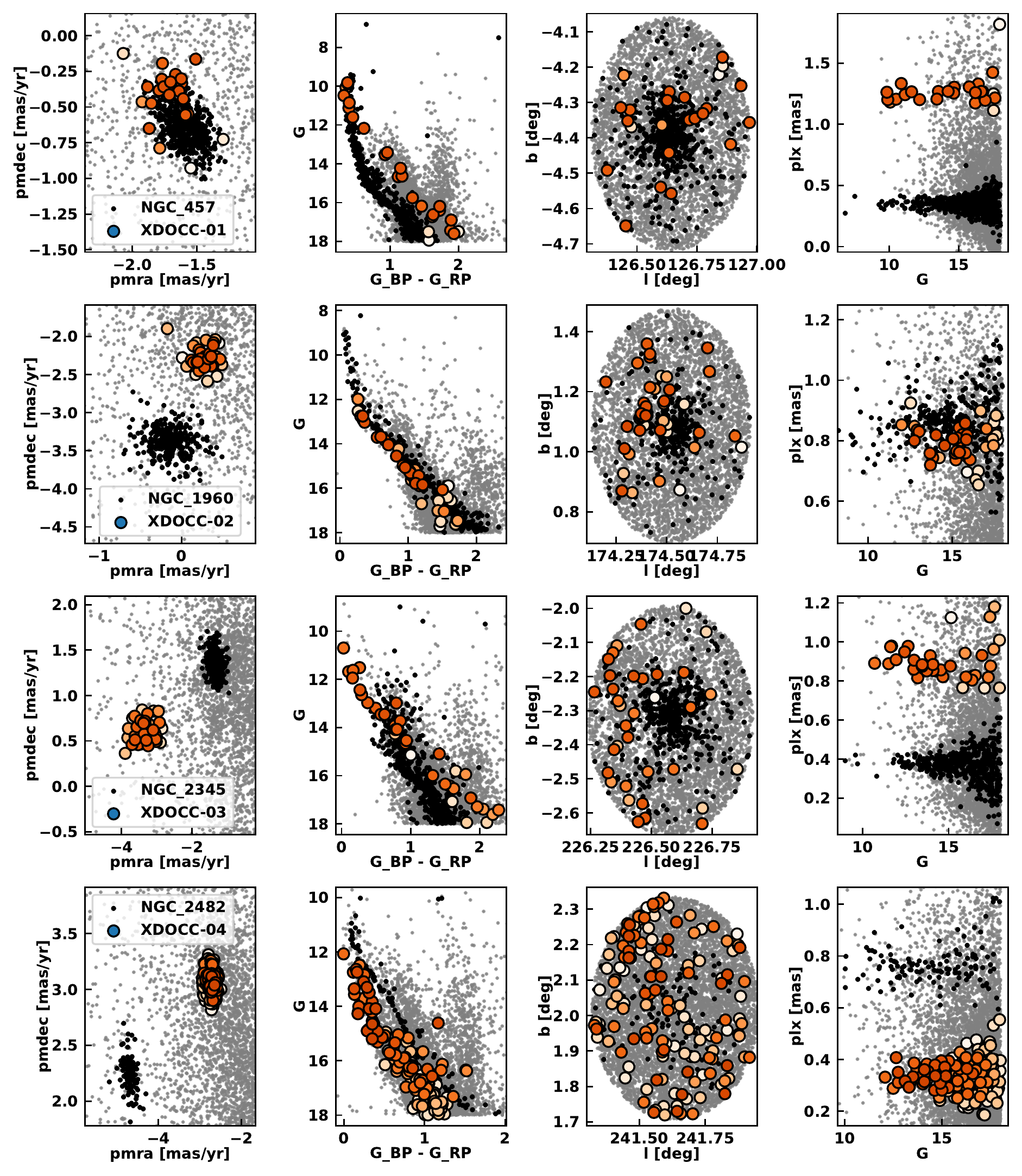}
\parbox{18cm}{\caption{Four plots for newly discovered XDOCC clusters [1-4]. Each cluster candidate gets a row in the figure. The first column of plots are proper motion vector plots. The second column are all Gaia DR2 \bp\ - \rp\ color and G-band magnitude plots. The third columns are the cluster galactic positions ({\it{l, b}}) in the sky. The fourth column are all parallax-magnitude plots. The XDOCC cluster is plotted as orange colored circles in all plots with the darker colors == higher membership probability. The targeted cluster found within the target field is plotted as black dots. The field is plotted in the background as small gray dots.  \label{fig:new_xdocc_clusters_page1}  }} 
% \end{center}
\end{figure*}

\begin{figure*}
% \epsscale{1.2}
% \begin{center}
\hspace{-0.75cm}
% \vspace{1cm}
\includegraphics[scale=0.99]{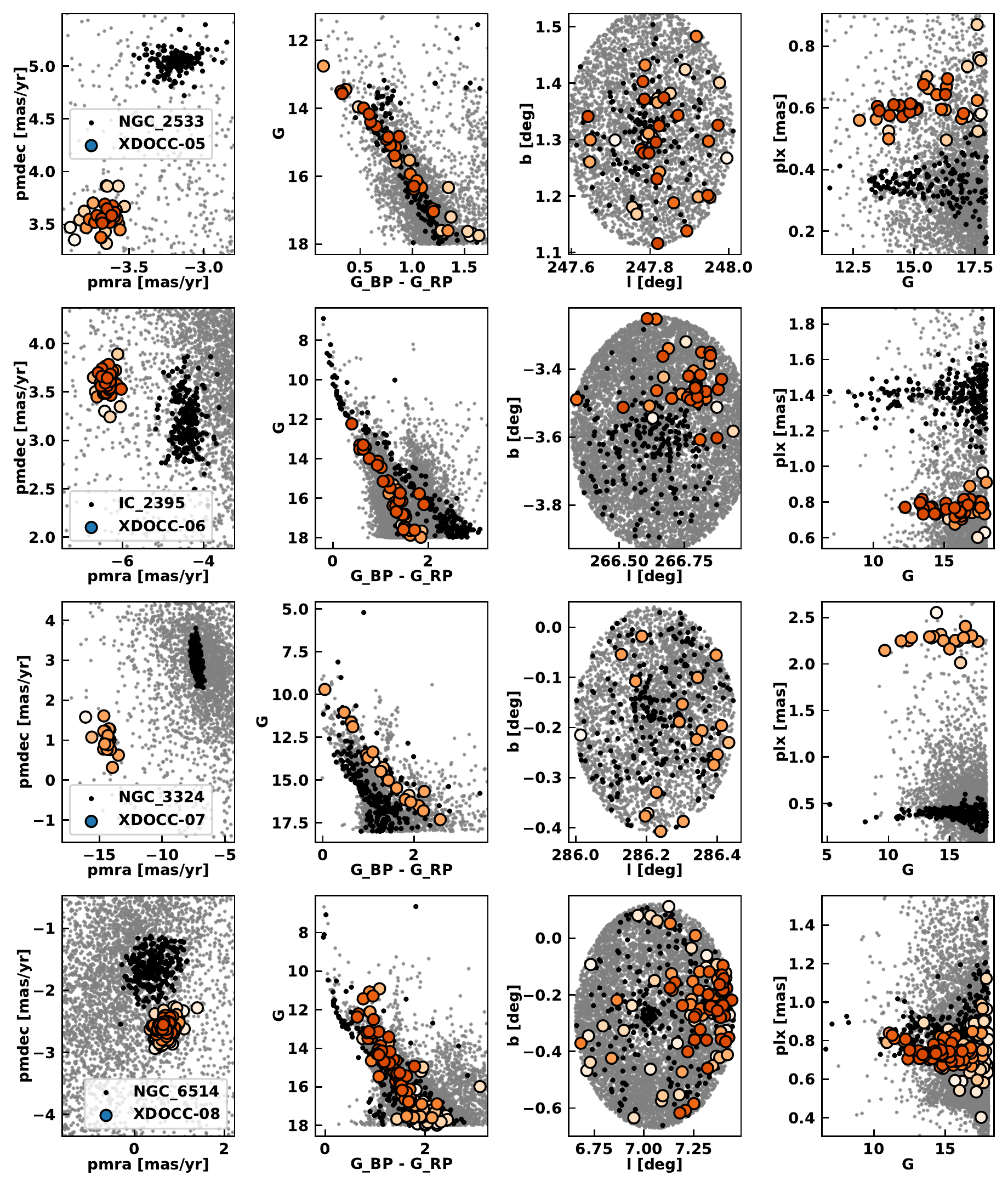}
\parbox{18cm}{\caption{Same as Figure \ref{fig:new_xdocc_clusters_page1}, but for XDOCC clusters [4-8]. \label{fig:new_xdocc_clusters_page2} }} 
% \end{center}
\end{figure*}

\begin{figure*}
% \epsscale{1.2}
% \begin{center}
\hspace{-0.75cm}
% \vspace{1cm}
\includegraphics[scale=0.99]{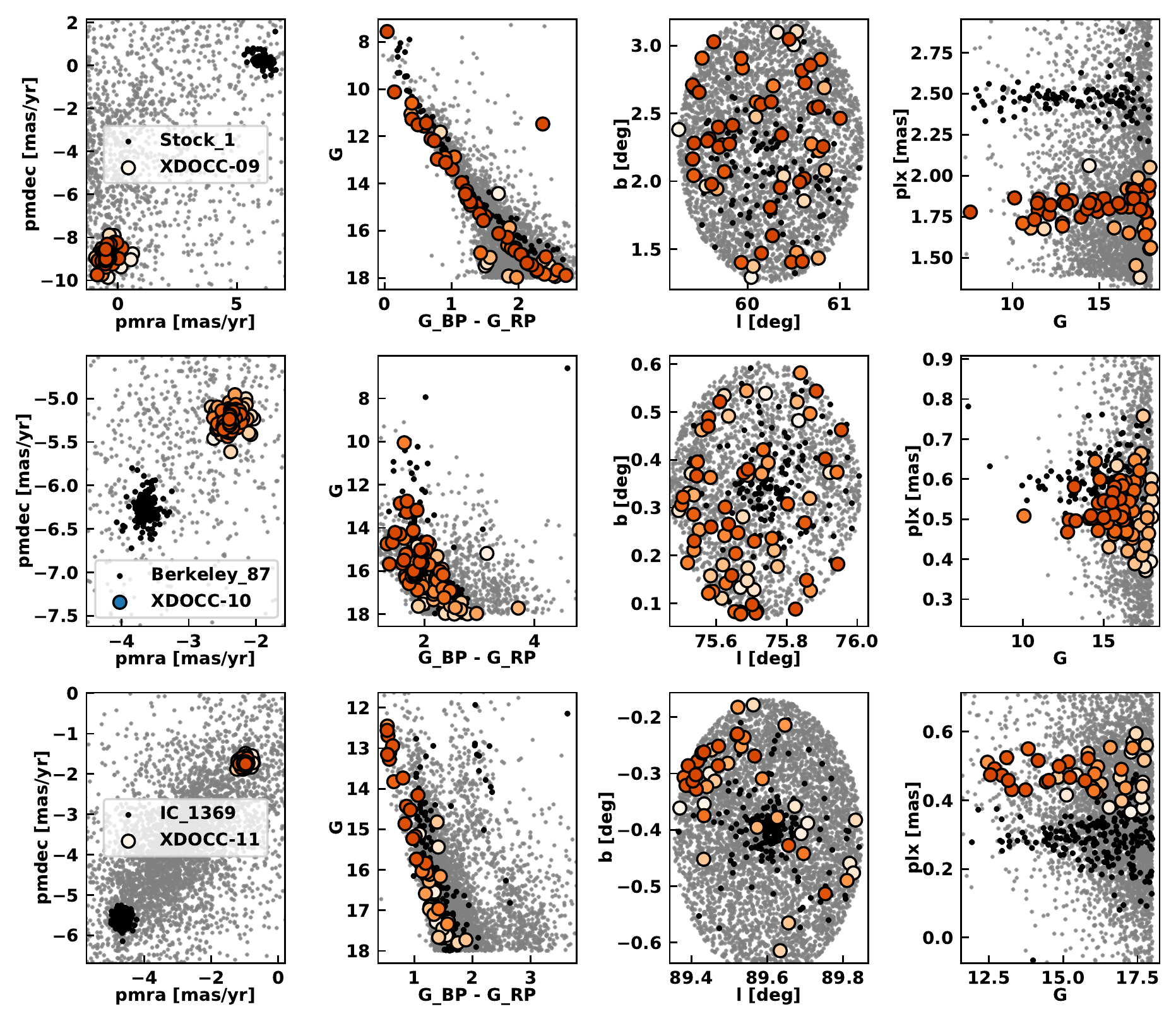}
\parbox{18cm}{\caption{Same as Figure \ref{fig:new_xdocc_clusters_page2}, but for XDOCC clusters [9-11]. \label{fig:new_xdocc_clusters_page3}  }} 
% \end{center}
\end{figure*}

\bibliography{output.bbl}
\bibliographystyle{aasjournal}



\end{document}